\documentclass[american,superscriptaddress,aps,prd,preprint,floatfix]{revtex4}
\usepackage[latin1]{inputenc}
\usepackage{float}
\usepackage{amsmath}
\usepackage{graphicx}
\usepackage{amssymb}
\usepackage{cancel}
\usepackage{epsfig}
\usepackage[dvips]{color}
\usepackage{multirow}
\usepackage{subfig}
\usepackage{psfrag}

\usepackage[bookmarks=true]{hyperref}
\usepackage{slashed}
\usepackage{lscape}

\usepackage[english]{babel}
\makeatother
\begin{document}

\preprint{Edinburgh 2013/14}
\preprint{CP3-Origins-2013-018 DNRF90}
\preprint{DIAS-2013-18}

\title{A note on Rome-Southampton Renormalization with Smeared Gauge Fields.}

\newcommand\edinb{SUPA, School of Physics, The University of Edinburgh,
                  Edinburgh EH9 3JZ, UK}
\newcommand\shojiaff{High Energy Accelerator Research Organization (KEK), Tsukuba 305-0801,
Japan and School of High Energy Accelerator Science, The Graduate University for
Advanced Studies (Sokendai), Tsukuba 305-0801, Japan}
\newcommand\rudyaff{$\text{CP}^3$-Origins and the Danish Institute for Advanced Study DIAS,
University of Southern Denmark, Campusvej 55, DK-5230 Odense M, Denmark.
}

\author{R. Arthur}     \affiliation{\rudyaff}
\author{P. A. Boyle}     \affiliation{\edinb}
\author{S. Hashimoto}     \affiliation{\shojiaff}
\author{R. Hudspith}     \affiliation{\edinb}

\date{\today}

\begin{abstract}

We have calculated continuum limit step scaling functions of bilinear and four-fermion operators
renormalized in a Rome-Southampton scheme using various smearing prescriptions
for the gauge field. Also, for the first time, we have calculated non-perturbative
anomalous dimensions of operators renormalized in a Rome-Southampton scheme.
The effect of such smearing first enters connected fermionic correlation functions via radiative corrections.
We use off-shell renormalisation as a probe,
and observe that the upper edge of the Rome-Southampton window is reduced by
link smearing.  This can be interpreted as arising due to 
the fermions decoupling from the high momentum gluons and we observe that 
the running of operators with the scale
at large lattice momenta shows enhanced lattice artefacts.
We find that the effect is greater for HEX smearing than for Stout
smearing, but that in both cases additional care must be taken when using 
off-shell renormalisation with smeared gauge
fields compared to thin link simulations.


\end{abstract}

\maketitle

\section{Introduction}

Link smearing is a popular method in lattice QCD, and is simple to 
combine with any fermion action. Certain smearing types are differentiable, {\it e.g.}  
Stout \cite{Morningstar:2003gk} and HEX \cite{Capitani:2006ni},
and can therefore be used in the calculation of the fermionic force term in hybrid Monte Carlo.
For staggered fermions, smearing has been motivated as
suppressing taste breaking by reducing gluon exchanges of momenta $\sim\pi/a$ \cite{Orginos:1999cr}.
Smearing improves the behaviour of lattice perturbation theory by 
making the tadpole contribution vanishingly small \cite{Bernard:1999kc}.
Smearing with Wilson fermions
has been very successful in reducing chiral symmetry breaking
errors by suppressing dislocations that lead to eigenvalues below
$m^2$ of $H_W^2$, ($H_W$ is the hermitian Wilson-Dirac operator $H_W=\gamma_5(D_W+m)$
with $D_W$ the Wilson-Dirac operator and $m$ the quark mass),
\cite{Hasenfratz:2007rf}. This allows smaller masses to be simulated on
coarser lattices without exceptional configurations.

Small eigenvalues of $\gamma_5(D_W-M)$ \footnote{$-M$ is a large negative mass term called the domain-wall height}
are responsible for residual chiral symmetry breaking
in domain wall fermions at finite $L_s$, the size of the fifth dimension \cite{Antonio:2008zz}.
This manifests itself in simulations as the residual mass $m_{res}$.
By reducing the number of modes in the region where the domain wall fermion's `$\text{tanh}$'
approximation to the sign function is most inaccurate it is hoped that one could obtain small $m_{res}$
without increasing $L_s$. 

By suppressing the low modes we are simultaneously
improving the condition number of the Dirac operator. Thus smearing should also
help by speeding up matrix inversions, as has already been observed for the 
overlap operator \cite{Hasenfratz:2007rf}.

One might worry that decoupling the fermion sector from the 
QCD gluodynamics at a scale below the lattice cut off 
might carry some penalty,
and it is prudent to carefully check the effects of this.
For some time it has been known that quantities like the static potential
at short distances \cite{Hasenfratz:2001hp} are strongly
distorted by smearing. In the case of the static potential
the distortion occurs for distances $r$ with $\frac{r}{a} < 2$, where lattice artefacts are large anyway.
It is also known that link smearing drives the renormalization constants towards their tree level values\cite{Kurth:2010yk}. 

The Rome-Southampton method for non-perturbative
renormalization (NPR) \cite{Martinelli:1994ty} is widely used. Steady improvements
to the original method: momentum sources \cite{Gockeler:1998ye}, non-exceptional momenta \cite{Aoki:2007xm}
and twisted boundary conditions \cite{Arthur:2010ht}, have led to a
precise method for renormalizing in lattice calculations (see \cite{Aoki:2010pe} for example).
The remaining systematic error is dominated by a perturbative
matching between the intermediate scheme, such as the RI/MOM, and the
$\overline{MS}$ scheme, which is done at a scale accessible by current lattice
calculations. The perturbative conversion error can be reduced by increasing the matching scale.
The available lattice momentum scale, $\frac{\pi}{a}$, can be of the order $3 \text{ GeV}$
to $5 \text{ GeV}$ which is high enough to reduce the matching error significantly.
A method has been developed \cite{Arthur:2010ht} (see also
\cite{Durr:2010aw}) that uses multiple lattices to stay below the lattice cut-off at all stages of
the calculation and calculate non-perturbative renormalization constants at
high energies where the effects of low energy QCD are minimised and perturbation
theory gives a good description of the running, which should make it possible to
reduce the error even further in the future. Link smearing and non-perturbative renormalization have 
often been used together to predict quantities of great
phenomenological importance, {\it e.g.}
in recent calculations of the kaon bag parameter $B_K$ \cite{Aubin:2009jh, Durr:2011ap}.

Rome-Southampton vertex functions may also be a case where the 
short distance properties are of 
crucial importance, especially since momentum scales 
close to the lattice cut-off are important.
In this paper we perform a systematic study of the effects 
of smearing on the high momentum behaviour of NPR
vertex functions and compare to the unsmeared ``thin link'' case. 
The continuum limit is taken at all stages assuming an 
$O(a^2)$ scaling violation and results are compared in the 
continuum limit where universality should be satisfied.

We aim to check how link smearing would affect the
program of Rome-Southampton renormalization that has been
used for kaon physics simulations in the RBC/UKQCD collaboration. 
The results should be of interest for non-perturbative renormalization with any action
since we will extrapolate smeared and unsmeared data to the continuum. The crucial
point in all that follows is that the continuum limit results should be independent
of the details of the intermediate lattice calculations, in particular if smeared links
have been used or not.

In section \ref{sec:fq} we calculate the smearing form factor and the associated smearing radius non-perturbatively.
Section \ref{sec:RRLMethods} describes our method of computing step scaling functions
and anomalous dimensions from NPR vertex functions and extrapolating to vanishing lattice spacing. The calculation
of anomalous dimensions in a Rome-Southampton scheme is a new method introduced in this paper.
Section \ref{sec:Results} shows NPR vertex functions calculated using different smearing prescriptions,
and derives step scaling functions and anomalous dimensions from them. Finally
we conclude in section \ref{sec:Conclusions} and comment on the use of smeared links together with
Rome-Southampton renormalization.

In this work we have not generated gauge fields using a smeared action;
in order for direct comparison we calculate the relevant
vertex functions on the same gauge ensembles with different smearing
parameters. For the gauge ensembles we use those of the RBC/UKQCD collaboration,
\cite{Allton:2008pn, Aoki:2010dy}. They are generated using the Iwasaki gauge action
and thin-link domain-wall fermions at lattice spacings $a^{-1} = 1.73(3) \text{ GeV}$ on a $24^3 \times 64 \times 16$
volume and $a^{-1} = 2.28(3) \text{ GeV}$ on a $32^3 \times 64 \times 16$
volume, that we refer to as the coarse and fine lattices respectively. The bare quark masses available
on the coarse lattice are $am = {0.005,0.01,0.02}$ with $m_{res} = 0.00315(4)$
and on the fine lattice  $am = {0.004,0.006,0.008}$,
$m_{res} = 0.000666(8)$. 

\section{Effect of Smearing on the Gauge Field}\label{sec:fq}

The effect of smearing on the lattice
Feynman rules is to introduce a form factor in the 
quark-gluon vertex \cite{Bernard:1999kc}.
After applying $n$ steps of smearing to obtain the gauge link $V^{(n)}_\mu( x + a \hat{\mu} /2 )$ we can define a gauge field through
\begin{equation}
 V^{(n)}_\mu(x + a \hat{\mu} /2) = \exp(ia A^{(n)}_\mu(x + a \hat{\mu} /2)).
\end{equation}
At leading order, the smeared and unsmeared gauge fields are related by,
\begin{equation}\label{eqn:smdef}
A^{(1)}_\mu(x + a \hat{\mu} /2) = \sum_{y, \nu} h_{\mu \nu}(y + a \hat{\nu} /2) A^{(0)}_\nu(x + a \hat{\mu} /2+y + a \hat{\nu} /2).
\end{equation}
with the momentum space the form factor $h_{\mu \nu}$,
\begin{eqnarray}\label{eqn:hdef}
 h_{\mu \nu}(q) &=& f^{(1)}(q) \left(\delta(q) - \frac{q_\mu q_\nu}{q^2}\right) + \frac{q_\mu q_\nu}{q^2},
\end{eqnarray}
$q_\mu = \frac{2}{a} \sin\left( \frac{a q_\mu}{2} \right)$.
Applying the smearing transformation $n$ times changes $f^{(1)}(q)$ to $f^{(n)}(q) = (f^{(1)}(q) )^n$.

The form factor $f^{(n)}(q)$ is \cite{Capitani:2006ni},
\begin{equation}\label{eqn:fqdef}
f^{(n)}(q) = \left[ 1 - \frac{\alpha}{2(d-1)} q^2 \right]^n
\end{equation}
for APE smearing with parameter $\alpha$ in $d$ dimensions. Stout has the same form.
For HYP and HEX $\alpha q^2$ should be replaced by $\alpha_1 \Omega_{\mu \nu} p^2_\nu$,
where $\Omega_{\mu \nu} = \left( 1 + \alpha_2(1 + \alpha_3) \right) \delta_{\mu \nu}$ to lowest order in $q^2$.
For small $q^2$ we can write the term in brackets as an exponential
\begin{equation}\label{eqn:fq_exp}
f^{(n)}(q) = \exp \left( -\frac{n \alpha }{2 (d-1) } q^2 \right) + O( (q^2)^2 ),
\end{equation}
and identify the mean square radius, 
\begin{equation}
\langle r^2 \rangle = \frac{n \alpha}{(d-1)},
\end{equation}
as a measure of the space-time region affected by the smearing transformation, in most
practical simulations this radius is of the order one.
This new distance scale introduced by smearing could alter the high energy
behaviour of the theory. In particular, since NPR directly probes radiative loop corrections
it may be highly sensitive to this new scale.

We fix the gauge fields to the Landau gauge and then
smear them using different smearing types: Stout, HYP and HEX. 
HYP smearing is described in \cite{Hasenfratz:2001hp} and HEX in \cite{Capitani:2006ni}. 
Both use the same hypercubic
blocking transformation but HEX uses the STOUT projection to $SU(3)$ and HYP uses the APE projection.
HYP and HEX smearing require the specification of three parameters,
$(\alpha_1, \alpha_2, \alpha_3)$ which we choose as $\{0.75,0.60,0.30\}$ for HYP (this is
sometimes called HYP-1) and $\{0.95,0.76,0.38\}$ for HEX. We show results using one and two hits of HYP
smearing and two hits of HEX smearing. We also show three hits of Stout with parameter
$\alpha_1 = 0.1$, using the convention of \cite{Morningstar:2003gk}. 
Using Appendix A of \cite{Capitani:2006ni} we can therefore calculate the tree level 
values of $\langle r^2 \rangle$, which we give in the first row of Table \ref{tab:effective_r}.  The values in this
table are small, indicating the effect of smearing hardly extends beyond one lattice spacing.

It is possible to calculate the quantity $f^{(n)}(q)$ defined in equation
(\ref{eqn:fqdef}) non-perturbatively as follows. Due to the Landau gauge
fixing, the equation $q_\mu A^{(0)}_\mu = 0$ holds to high precision. From equation (\ref{eqn:hdef})
this means,
\begin{equation}
A^{(n)}_\mu(q) = f^{(n)}(q) A^{(0)}_\mu(q),
\end{equation}
and hence
\begin{equation}\label{eqn:deffq}
 f^{(n)}(q)  = \left( \frac{\textrm{Tr}[A^{(n)}_\mu(q) A^{(n)}_\mu(-q)]}{\textrm{Tr}[A^{(0)}_\mu(q) A^{(0)}_\mu(-q)]
}\right)^{1/2}.
\end{equation}
The lattice gluon field is defined by,
\begin{multline}
A_\mu(x + a \hat{\mu} /2) = \frac{1}{2ia}\Bigl( \left[ U_\mu(x + a \hat{\mu}/2) - U^\dagger_\mu(x + a \hat{\mu}/2) \right] \\
 - \frac{1}{N_c}\textrm{Tr}\left[ U_\mu(x + a \hat{\mu}/2) - U^\dagger_\mu(x + a \hat{\mu}/2) \right] I_{N_C \times N_C}
\Bigr),
\end{multline}
and in momentum space by,
\begin{equation}\label{Formalism:Eq:Ap_corrected}
A_{\mu}(q) =
e^{i\frac{q_\mu}{2}}\sum_{x}e^{iq\cdot{x}}A_{\mu}(x+a\hat{\mu}/2).
\end{equation}

\begin{figure}[htp]
  \begin{center}
    \subfloat[3 STOUT $f^{(3)}(q)$ weak field]{
\includegraphics*[angle=0,width=0.35\textwidth]{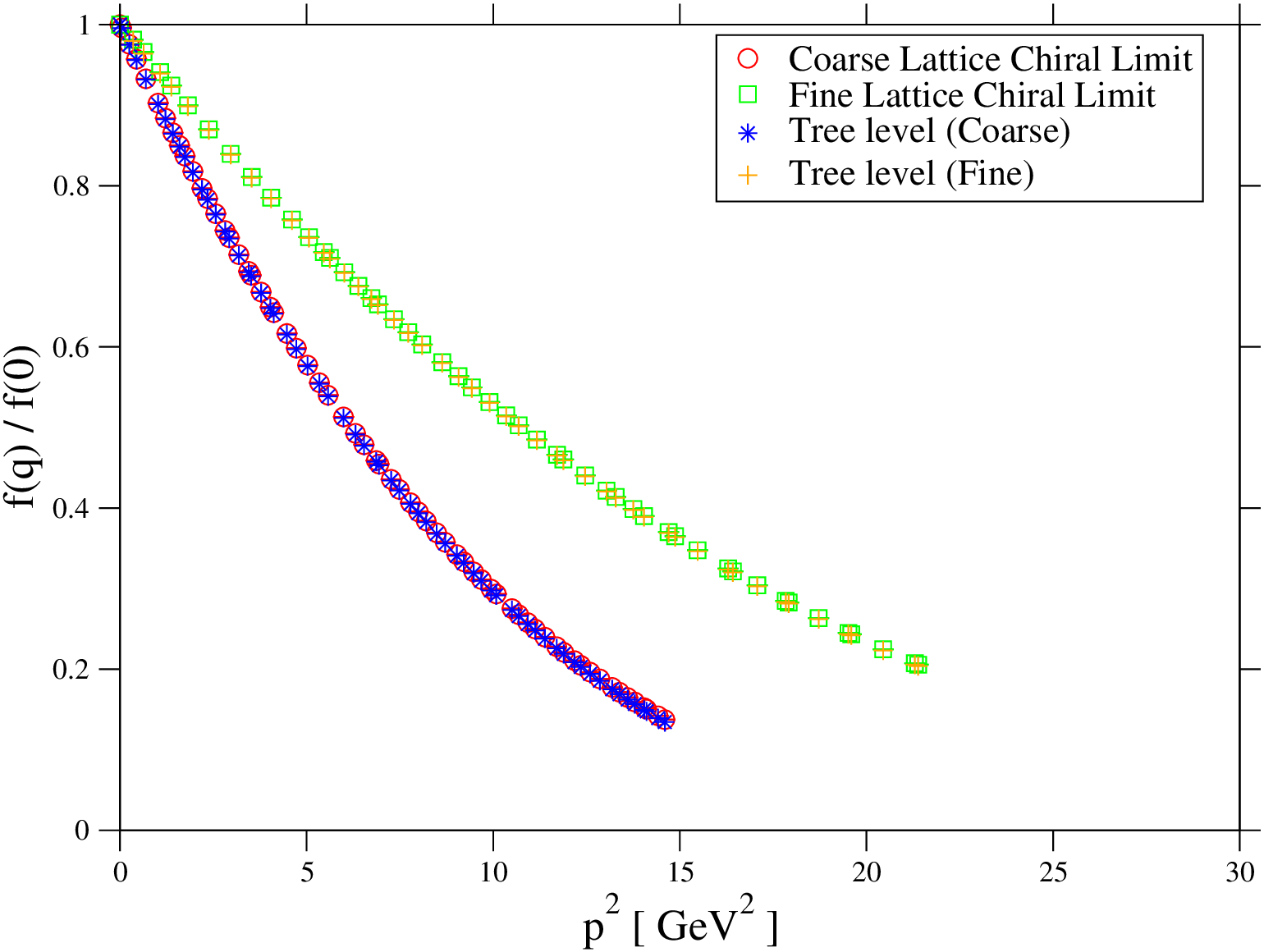}} 
    \subfloat[3 STOUT $f^{(3)}(q)$ full field]{
\includegraphics*[angle=0,width=0.35\textwidth]{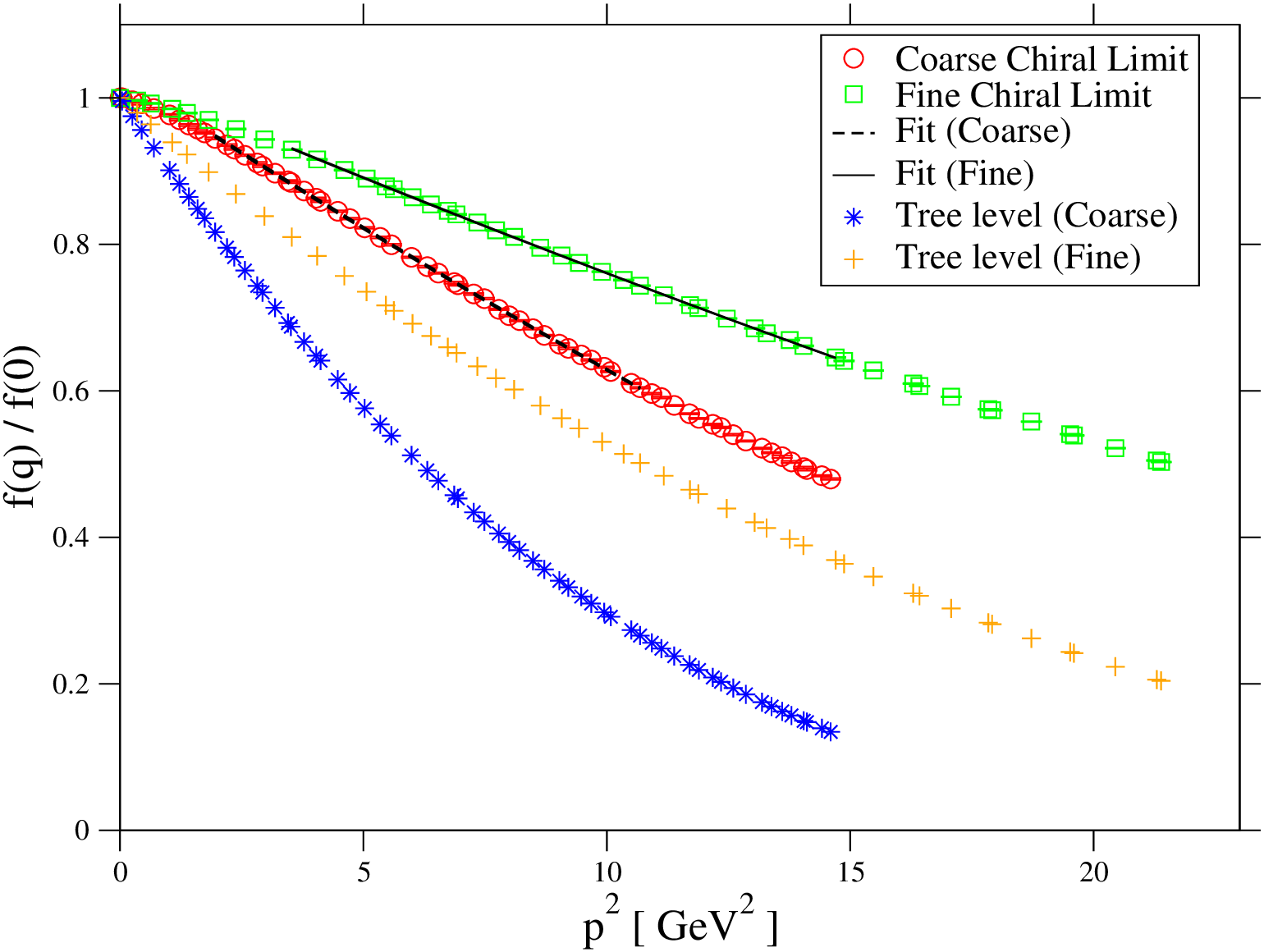}} \\
    \subfloat[1 HYP $f^{(1)}(q)$ weak field]{
\includegraphics*[angle=0,width=0.35\textwidth]{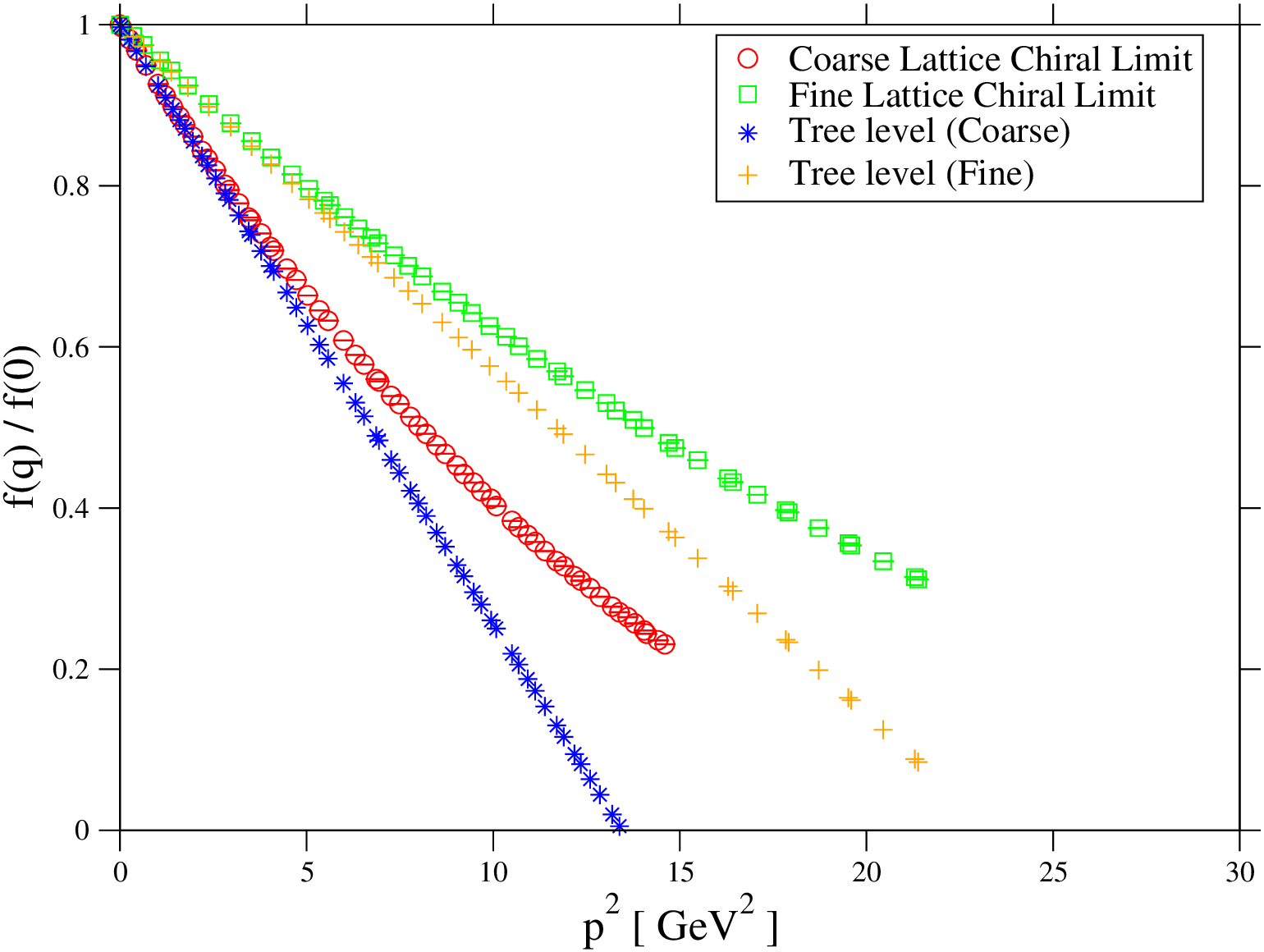}} 
    \subfloat[1 HYP $f^{(1)}(q)$ full field]{
\includegraphics*[angle=0,width=0.35\textwidth]{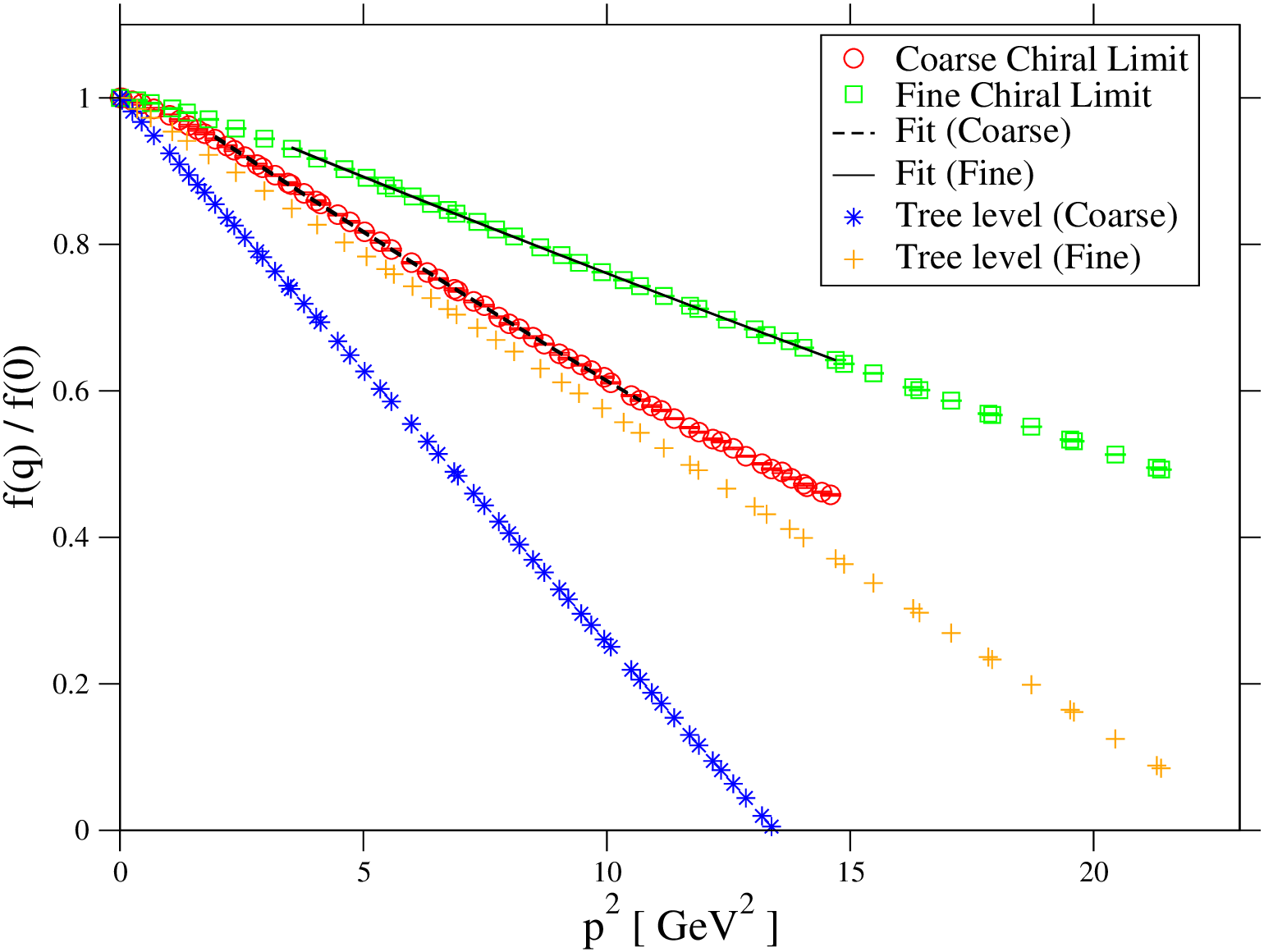}} \\
  \end{center}
  \caption{$f^{(n)}(q)$ measured on the coarse (red) and fine (green) ensembles for three hits of Stout, (a) and (b),
and one of HYP, (c) and (d). On the left are results using a weak field as a cross check for results on the right using the full field.
Perturbative predictions in blue and orange. Fits are to a quadratic in $q^2$ over the indicated range.
  }
  \label{fig:fq_plotsa}
\end{figure}

\begin{figure}[htp]
  \begin{center}
    \subfloat[2 HYP $f^{(2)}(q)$ weak field]{
\includegraphics*[angle=0,width=0.35\textwidth]{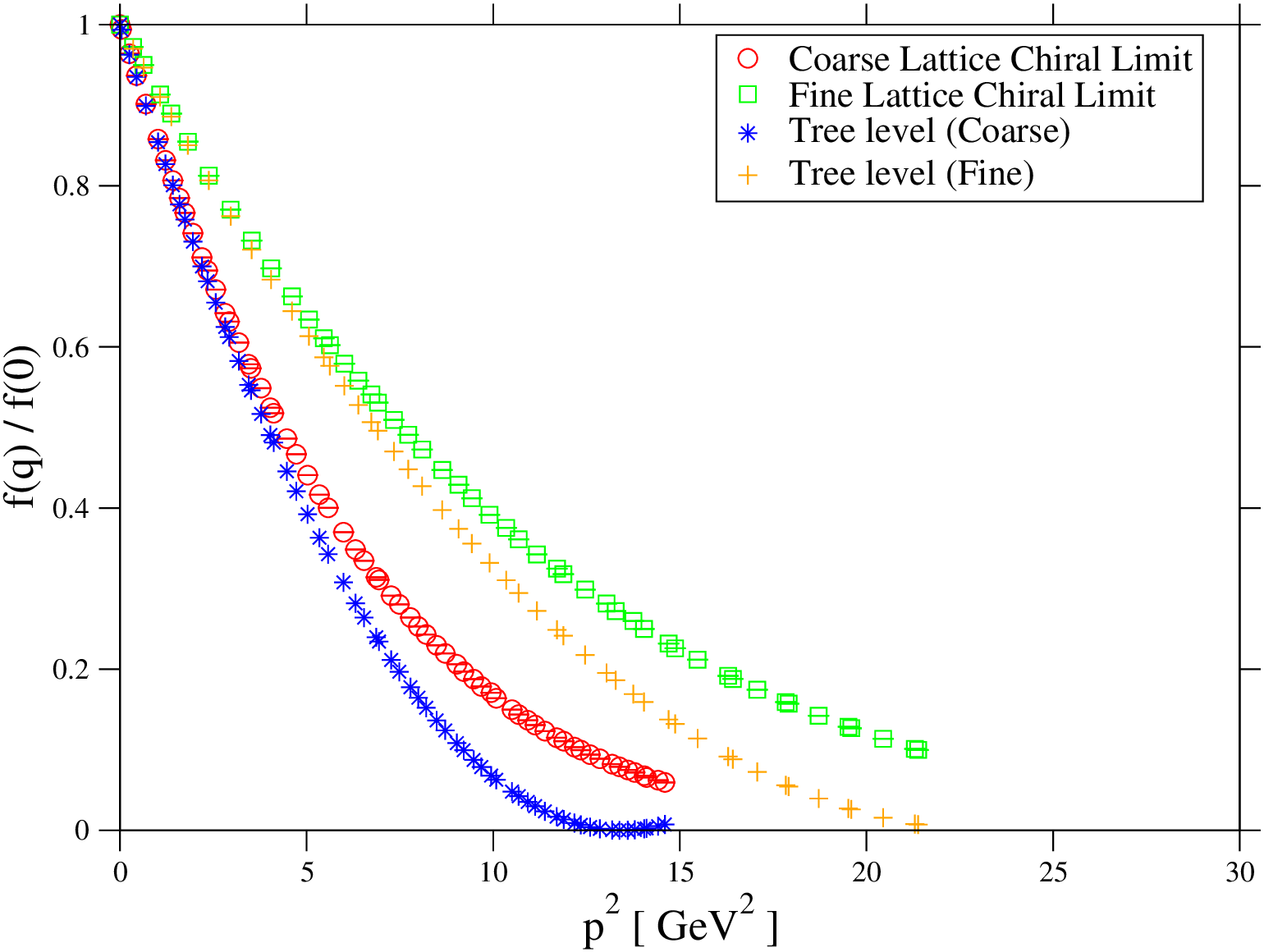}} 
    \subfloat[2 HYP $f^{(2)}(q)$ full field]{
\includegraphics*[angle=0,width=0.35\textwidth]{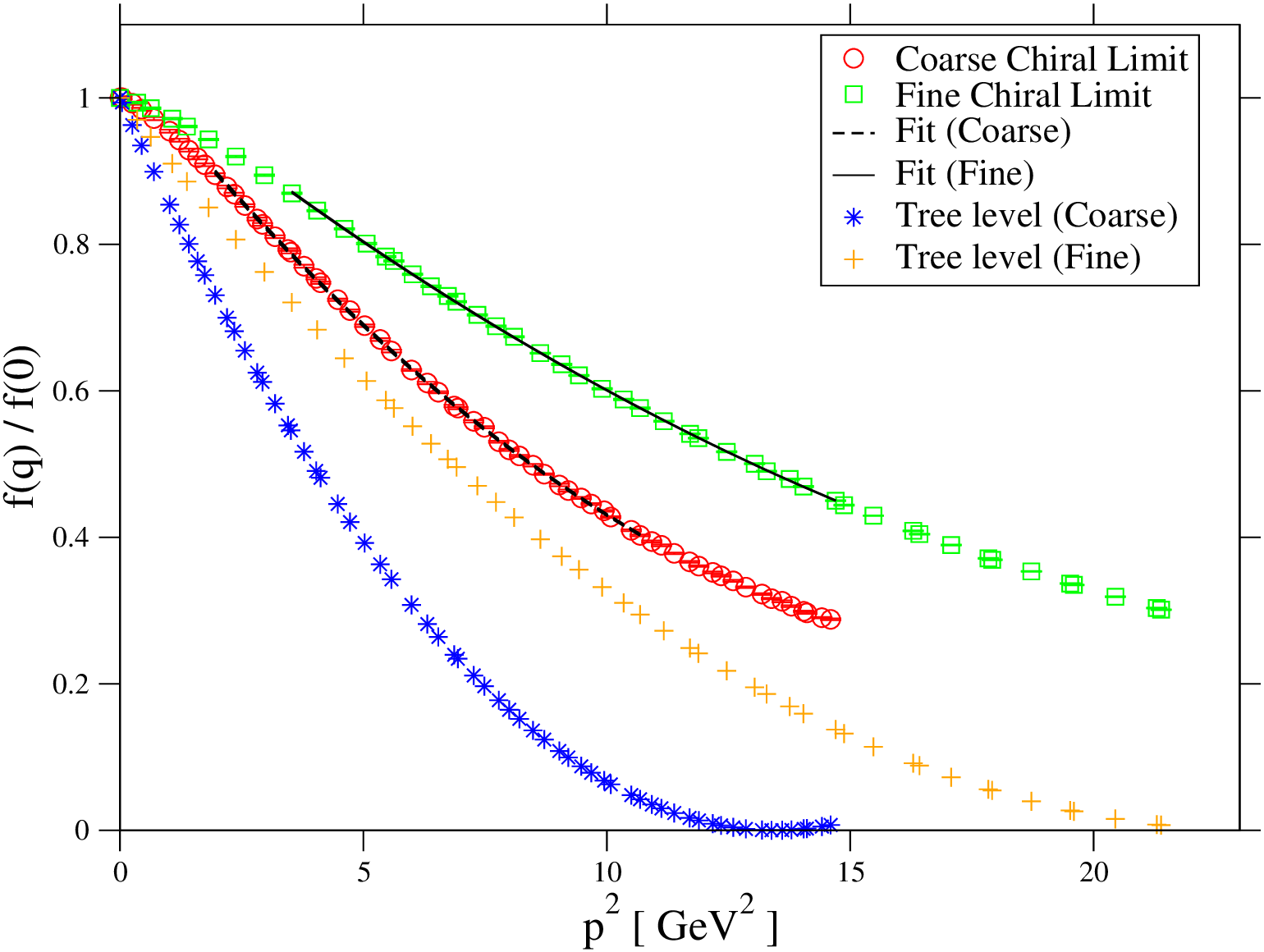}} \\
    \subfloat[2 HEX $f^{(2)}(q)$ weak field]{
\includegraphics*[angle=0,width=0.35\textwidth]{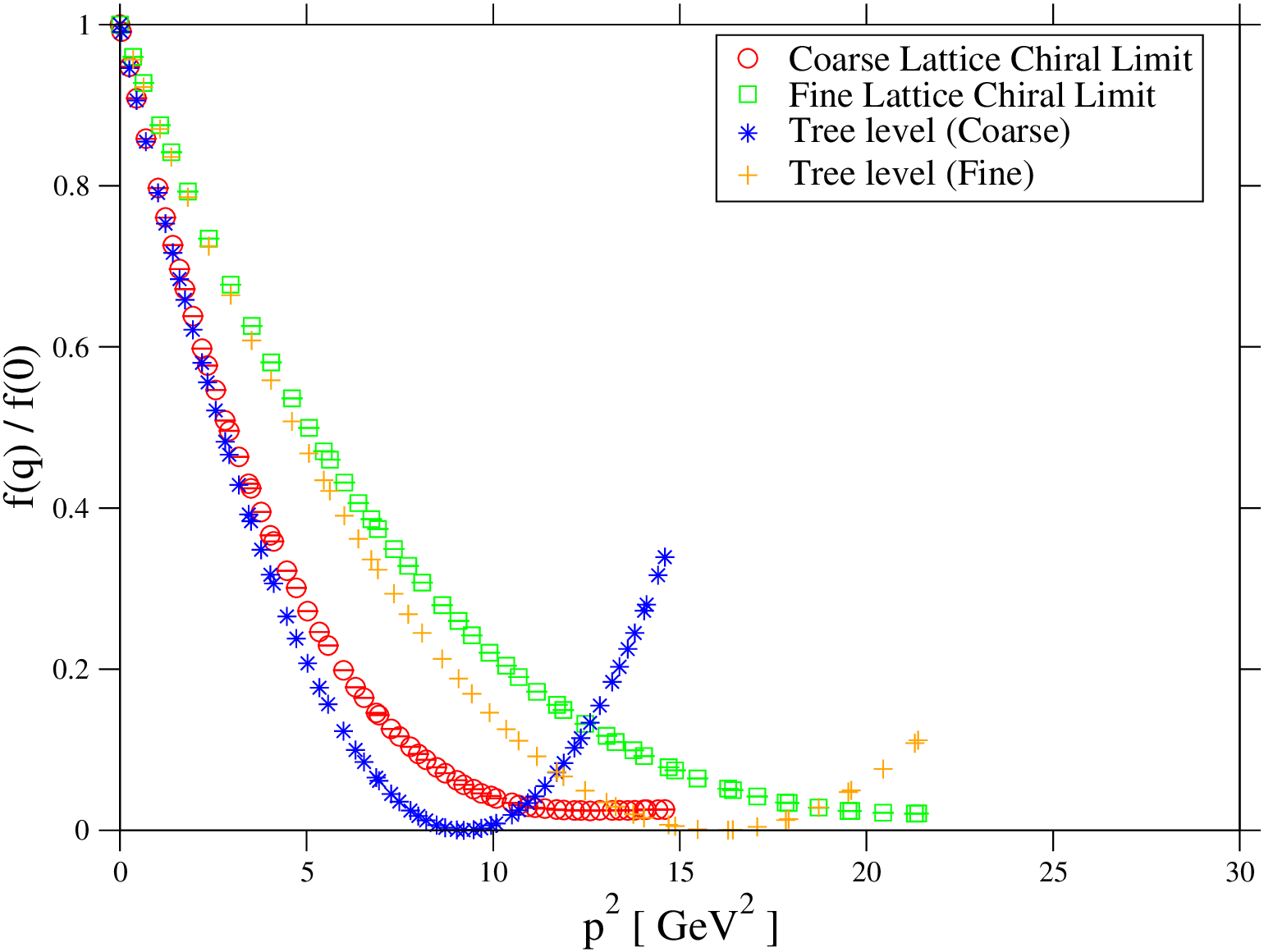}} 
    \subfloat[2 HEX $f^{(2)}(q)$ full field]{
\includegraphics*[angle=0,width=0.35\textwidth]{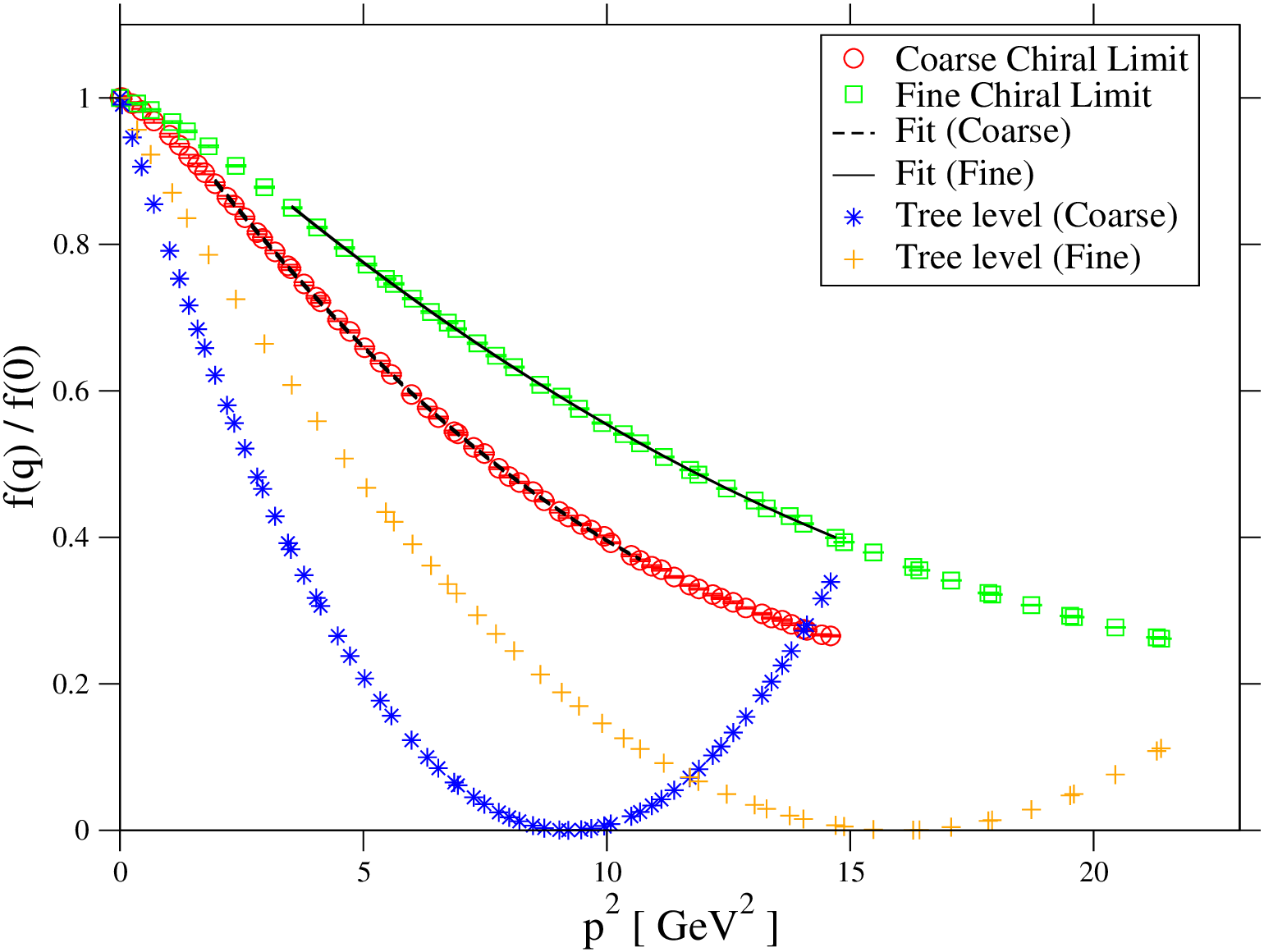}} \\
  \end{center}
  \caption{$f^{(n)}(q)$ measured on the coarse (red) and fine (green) ensembles for two hits of HYP, (a) and (b),
and two of HEX, (c) and (d). On the left are results using a weak field as a cross check for results on the right using the full field.
Perturbative predictions in blue and orange. Fits are to a quadratic in $q^2$ over the indicated range.
  }
  \label{fig:fq_plotsb}
\end{figure}


This can be used in equation (\ref{eqn:deffq}) and $f\left(q \right)$ calculated for different levels of smearing. 
In the left column of Figures \ref{fig:fq_plotsa} and \ref{fig:fq_plotsb}, (a cylinder cut in momentum space \cite{Leinweber:1998im} of
width $\frac{2\pi}{L_{i}}$, $i$ is the spatial lattice index, has been used)
are the `weak field' results. This is introduced to check the analysis against perturbation theory.
Namely, we multiply every gauge field $A_\mu$ by $10^{-3}$, generate $U_\mu$ by exact exponentiation 
and then apply the smearing. For this artificially produced weak field
we find good agreement with the perturbative expectation, as shown in
the left column of Figures \ref{fig:fq_plotsa} and \ref{fig:fq_plotsb}. The `full field'
results, using the original, unreduced $U_\mu$, are shown on the right.

The identification of the quantity $\sqrt{\langle r^2 \rangle}$ as an effective smearing
radius assumes a Gaussian profile for $f^{(n)}(q)$. From the plots of $f^{(n)}(q)$ 
it is clear that it is not Gaussian and fitting the curves with
a Gaussian fails. We attempt to extract a value of $\langle r^2 \rangle$ in two ways: (a)
using the point where $f^{(n)}(q)$ is equal to $e^{-1/2}$ of its initial value to define a scale,
\begin{equation}\label{eqn:expdef}
\langle r^2 \rangle = \frac{1}{q^2}\Big{|}_{f^{(n)}(q)/f^{(n)}(0) = e^{-1/2}}
\end{equation} 
(b) assuming that the functional form of $f^{(n)}(q)$ is the same as at tree level, 
\begin{equation}\label{eqn:fitdef}
f^{(n)}(q) = \left( 1 - \frac{1}{2n} \langle r^2 \rangle q^2 \right)^n = 1 - \frac{1}{2} \langle r^2 \rangle q^2 + cq^4 + O(q^6),
\end{equation}
and fitting to a quadratic, then obtaining $\langle r^2 \rangle$ from the linear co-efficient.
The results of both of these strategies are given in Table \ref{tab:effective_r}.

These results are far from the perturbative picture, giving significantly smaller radii.
In order to make the identification of the perturbative function $f^{(n)}(q)$ in equation (\ref{eqn:hdef})
with equation (\ref{eqn:deffq}) we assume that equation (\ref{eqn:smdef})
is valid. However this is only the case up to leading order in $a$,
the smeared gauge fields do not even satisfy the Landau condition.
Our calculation of $f^{(n)}(q)$ is the first time this has been measured non-perturbatively
and gives unexpectedly small smearing radii. 

Having seen how smearing affects the
high momentum part of the gauge field we turn to the main focus of this paper:
how smearing affects NPR at high scales.

\begin{table}[hbt]
\begin{tabular}{|c|c|c|c|c|c|}
\hline
  \multicolumn{2}{|c|}{Smearing} & 3 Stout & 1 HYP & 2 HYP & 2 HEX \\ \hline
  \hline
  \multirow{2}{*}{$\sqrt{ \langle r^2 \rangle }$} & tree level & 0.667 & 0.775 & 0.943 & 1.139 \\ 
& coarse ensemble (exp) & 0.5307(4) & 0.5418(3) & 0.6847(5) & 0.7165(4)\\
& coarse ensemble (fit) & 0.509(2) & 0.514(2) &  0.705(1) &  0.746(2)\\
& fine ensemble (exp) & 0.5625(2) & 0.5663(2) & 0.7271(2) & 0.7730(3)\\    
& fine ensemble (fit) & 0.543(1) & 0.540(1) & 0.747(1) & 0.802(1) \\
\hline

\end{tabular}
\caption{Effective smearing radius in lattice units measured using equation (\ref{eqn:expdef}); (exp)
and equation (\ref{eqn:fitdef}); (fit). The disagreement between the two definitions arises because the curves are not really
Gaussian.}\label{tab:effective_r}
\end{table}

\section{Renormalization and Running of Lattice Operators}\label{sec:RRLMethods}

Following \cite{Arthur:2010ht} we calculate step scaling functions in the continuum limit 
for a variety of operators. After obtaining the
Fourier transformed propagator $ S(z,p)$ by solving
\begin{equation}\label{eqn:momsrc}
 \sum_{z} D(x,z) S(z,p) = e^{i p x},
\end{equation}
we multiply by a phase to get $S'(z,p) = e^{-ipz} S(z,p)$ and calculate the 
Green function for the bilinear operator $O_\Gamma(z) = \bar{q}(z) \Gamma q(z)$
\begin{equation}
 \langle G_{\cal_O} (p_1,p_2) \rangle = \langle \frac{1}{V} \sum_{z} \gamma_5 (S'(z,p_1))^\dagger \gamma_5 \Gamma S'(z,p_2) \rangle. 
\end{equation}
We use the inverse propagator, $S^{-1}(p)$, to make the amputated vertex function,
\begin{equation}
 \Pi_{ {\cal O} } (p) =  \langle S^{-1}(p_1) \rangle \langle G_{ {\cal_O} } (p_1,p_2) \rangle \langle S^{-1}(p_2) \rangle .
\end{equation}
For details on the procedure for the renormalization of $B_K$ see \cite{Aoki:2010pe}.
Only non-exceptional momenta; $p_\mu$, $q_\mu$, $r_\mu$ where $r_\mu = p_\mu - q_\mu$ 
and $p^2 = q^2 = (p - q)^2$ are used.
To access momenta that are not Fourier modes we solve equation (\ref{eqn:momsrc}) 
using twisted boundary conditions.
The momenta in the direction $\mu$ are then modified such that $p_\mu$ becomes
$p_\mu + \frac{\pi \theta_\mu}{L_\mu}$, 
where $L_\mu$ is the extent
of the lattice in that direction and $\theta_\mu$ the $\mu$ component of a vector parallel $p$. 
We keep the directions of $p_\mu$,
$q_\mu$ and $r_\mu$ fixed and use different twists ($\theta$s) to vary their magnitudes.
In this way we avoid ${\cal H}(4)$ breaking artefacts \cite{Arthur:2010ht}.

We define the normalized trace of
the amputated vertex function with a projector $\mathbf{P}_{{\cal_O}}$ as,
\begin{equation}
 \Lambda_{{\cal_O}} (\mu, a, m ) = \frac{1}{12} Tr \left[ \mathbf{P}_{{\cal_O}} \Pi_{{\cal_O}} (p,a,m) \right]_{p^2 = \mu^2},
\end{equation}
where the $\mathbf{P}_{{\cal_O}}$ is chosen to project out a desired
gamma matrix structure and to give a normalization such that $\Lambda_{\cal O}$ is 
unity at tree level. The trace is over spinor and color degrees of freedom.
The renormalized amputated vertex functions of bilinears are related to the bare ones by
\begin{equation}
 \Lambda_{{\cal_O}, R}(\mu, a, m ) = \frac{Z_{{\cal_O}}(\mu, a, m ) }{Z_q}  \Lambda_{{\cal_O}} (\mu, a, m ),
\end{equation}
the $Z_q$ factor comes from the inverse propagators required for the external leg amputation. 
The Rome-Southampton renormalization condition on the projected vertex functions is,
\begin{equation}\label{eqn:vertexrenorm}
 \lim_{m \to 0} \Lambda_{{\cal_O} R} (\mu, a, m ) = 1.
\end{equation}

To look at the running of each operator individually we have to remove the factors of
$Z_q$ implicit in $\Lambda_{{\cal_O}}$. Dividing by the axial vertex $\Lambda_{A}$
\begin{equation}
 R_{ {\cal O} } (\mu, a, m) = \frac{\Lambda_{A}(\mu, a, m) }{\Lambda_{\cal O}(\mu, a, m)}
= \frac{Z_{ {\cal O} }(\mu, a, m) }{Z_A}
\end{equation}
accomplishes this since $Z_A$ is fixed by current conservation and does not run with $\mu$.
These quantities are extrapolated to the chiral limit to give the renormalization constant,
\begin{equation}
 Z_{ {\cal O} }(\mu, a) = Z_A \lim_{m\to 0} R_{\cal O}(\mu,a,m).
\end{equation}

We introduce a ratio of the $R_{\cal O}$s at different $\mu$,
\begin{equation}
\Sigma_{\cal O}(\mu, s \mu,a) = \lim_{m\to 0}
                        \frac{R_{\cal O}(s \mu,a,m)}{
                             R_{\cal O}(\mu,a,m)},
\end{equation}
which has,
\begin{equation}\label{eq:sig}
\sigma_{\cal O}(\mu, s \mu) = \lim_{a\to 0}\Sigma_{\cal O}( \mu, s \mu, a) = \frac{Z_{\cal O}(s\mu)}{Z_{\cal O}(\mu)}
\end{equation}
as its continuum limit and is called the step scaling function. This is the factor required to change the scale from
$\mu$ to $s \mu$. Because we take the continuum limit the final step scaling functions should be action
independent. To obtain the mass step scaling function we use the pseudoscalar bilinear operator
${\cal O} = \bar{q} \gamma_5 q$. For the quark field renormalization we use the vector bilinear ${\cal O}
 = \bar{q} \gamma_\mu q$
and for the tensor current we use ${\cal O} = \bar{q} \sigma_{\mu \nu} q$. The renormalization
of $B_K$ requires the VV+AA four quark operator $\left( \bar{q} \gamma_\mu q \right) \left( \bar{q} \gamma_\mu q \right) + 
\left( \bar{q} \gamma_5 \gamma_\mu q \right) \left( \bar{q} \gamma_5 \gamma_\mu q \right). $

In order to cleanly separate high and low energy behaviour it is useful to have a function that depends on only
one scale. This is given by the anomalous dimension $\gamma_{\cal O}$, 
\begin{equation}
\sigma_{\cal O}( \mu, s \mu ) = \exp\left( \int_{\alpha_1}^{\alpha_2} \frac{\gamma_{\cal O} (\alpha)}{ \beta(\alpha)} d\alpha \right) \\
= \exp\left( \int_{\mu_1}^{\mu_2} \frac{\gamma_{\cal O} (\mu)}{\mu} d\mu \right)
\end{equation}
using the chain rule. With $\mu_1 = \mu$ and $\mu_2 = (1+\epsilon) \mu$ for small $\epsilon$ the right hand side
becomes,
\begin{equation}\label{eqn:ad}
\sigma_{\cal O}\left( \mu, (1 + \epsilon) \mu \right) \approx \exp \left(\epsilon \gamma_{\cal O} (\mu) \right) \\
\rightarrow \gamma_{\cal O}(\mu) \approx \ln \left( \sigma_{\cal O}( \mu, (1 + \epsilon) \mu )  \right) / \epsilon.
\end{equation}
As we will see, the vertex function data, and hence the step scaling functions derived from it, are very smooth
functions of the scale which means they can be differentiated as above without difficulty.

\section{Results}\label{sec:Results}

\subsection{Vertex Functions}
\begin{figure}[htp]
  \begin{center}
    \subfloat[$\Lambda_{\cal A}$, $RI-SMOM$]{
\includegraphics*[angle=0,width=0.35\textwidth]{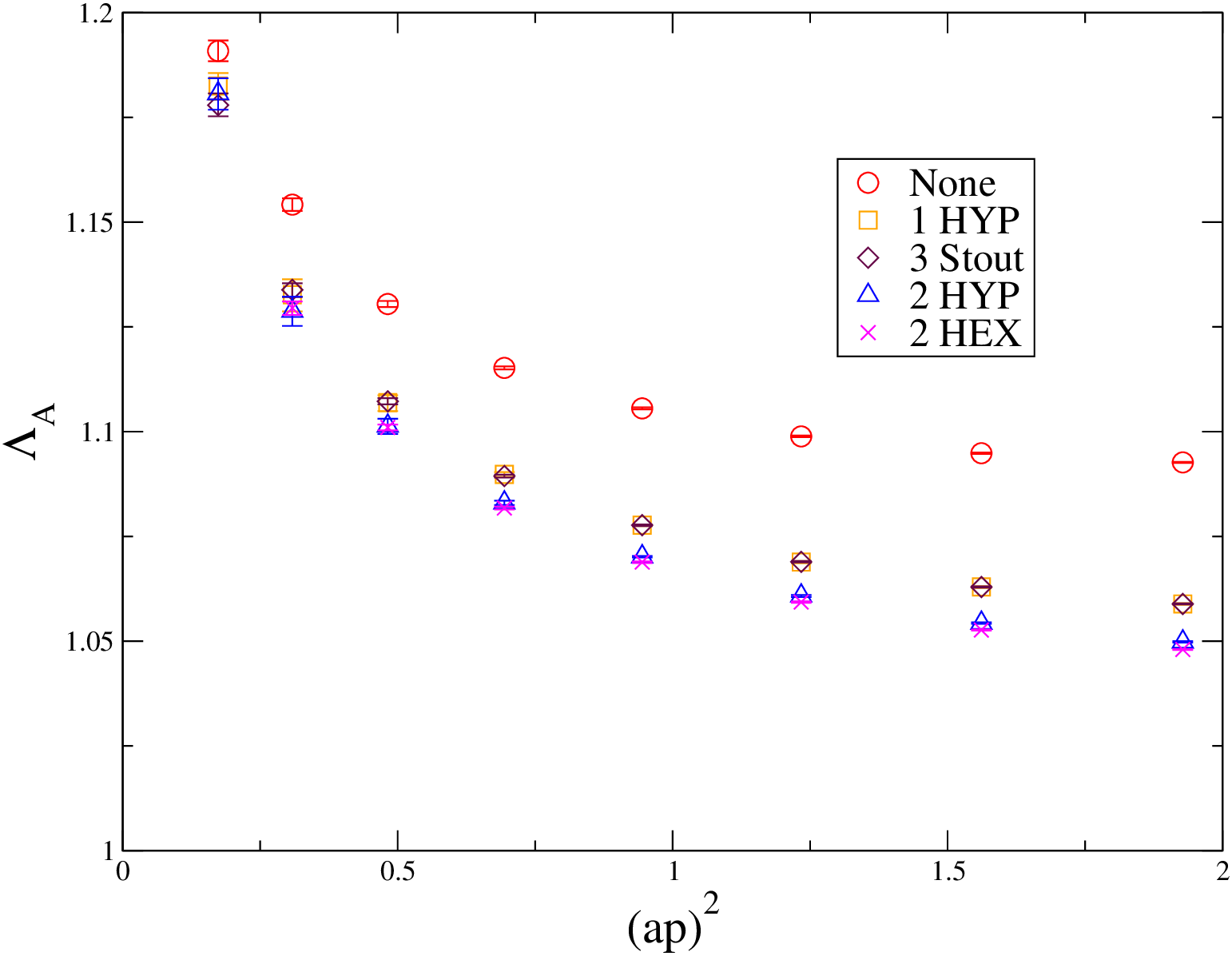}} 
    \subfloat[$\Lambda_{\cal V}$, $RI-SMOM$]{
\includegraphics*[angle=0,width=0.35\textwidth]{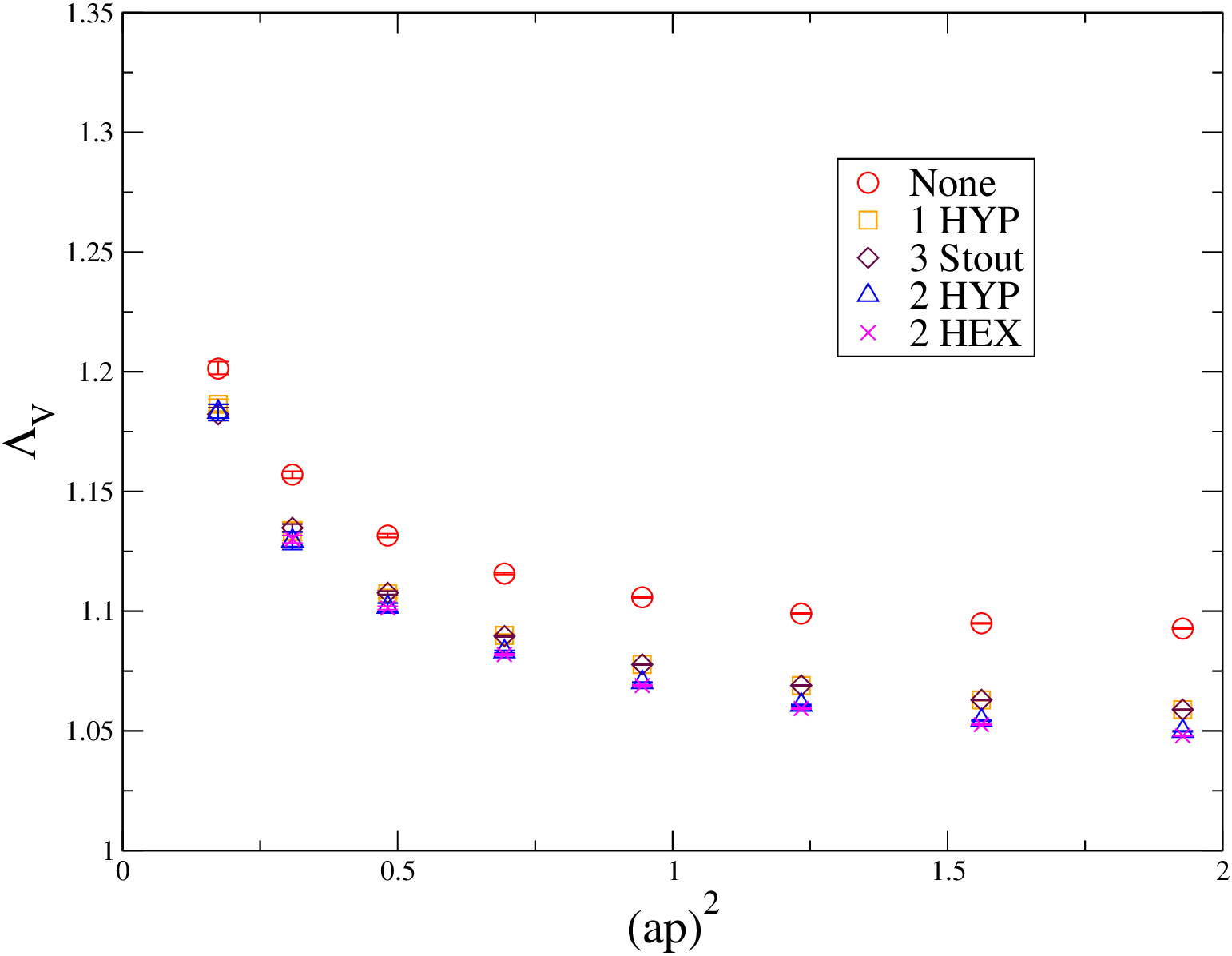}} \\
    \subfloat[$\Lambda_{\cal P}$, $RI-SMOM$]{
\includegraphics*[angle=0,width=0.35\textwidth]{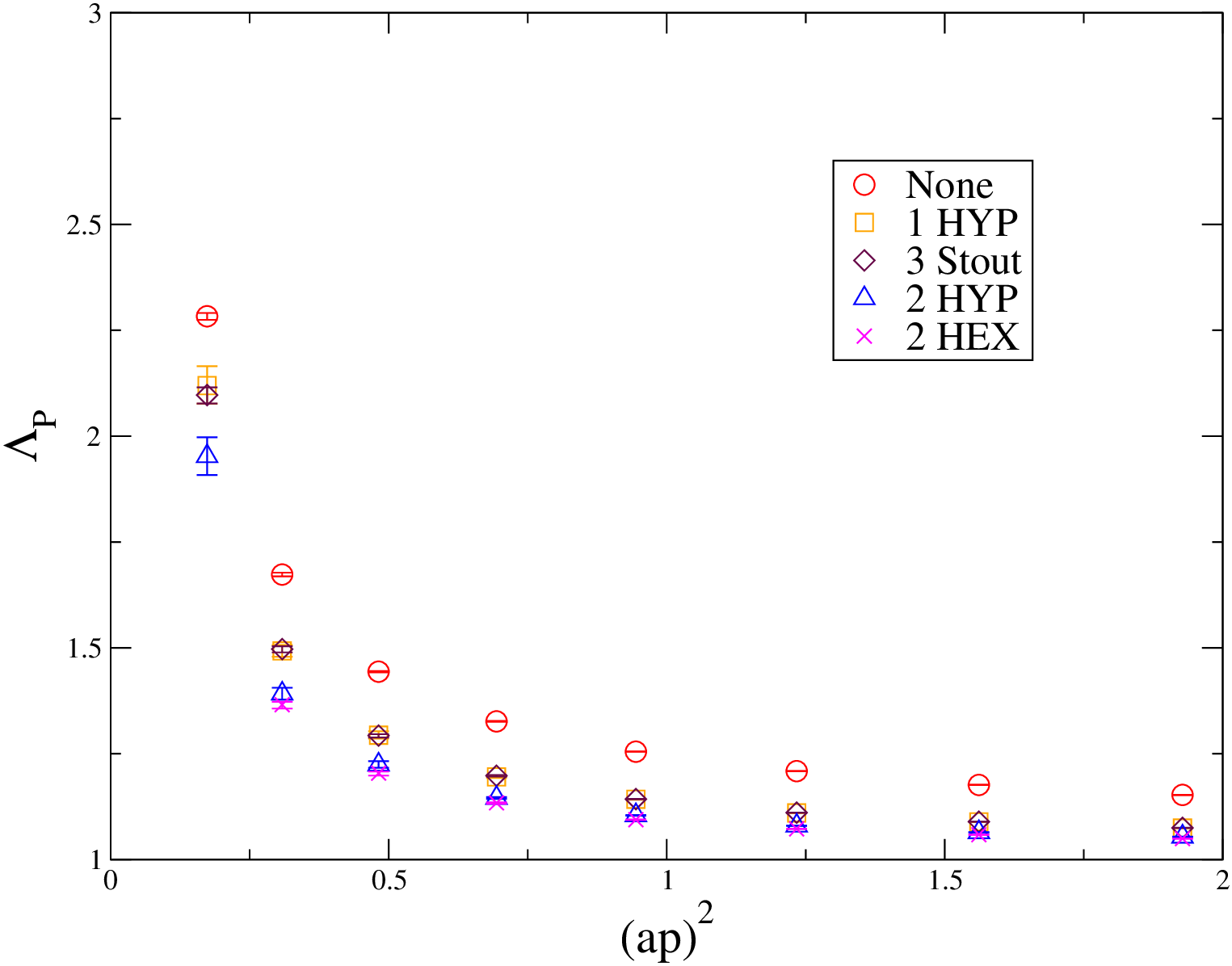}} 
    \subfloat[$\Lambda_{\cal S}$, $RI-SMOM$]{
\includegraphics*[angle=0,width=0.35\textwidth]{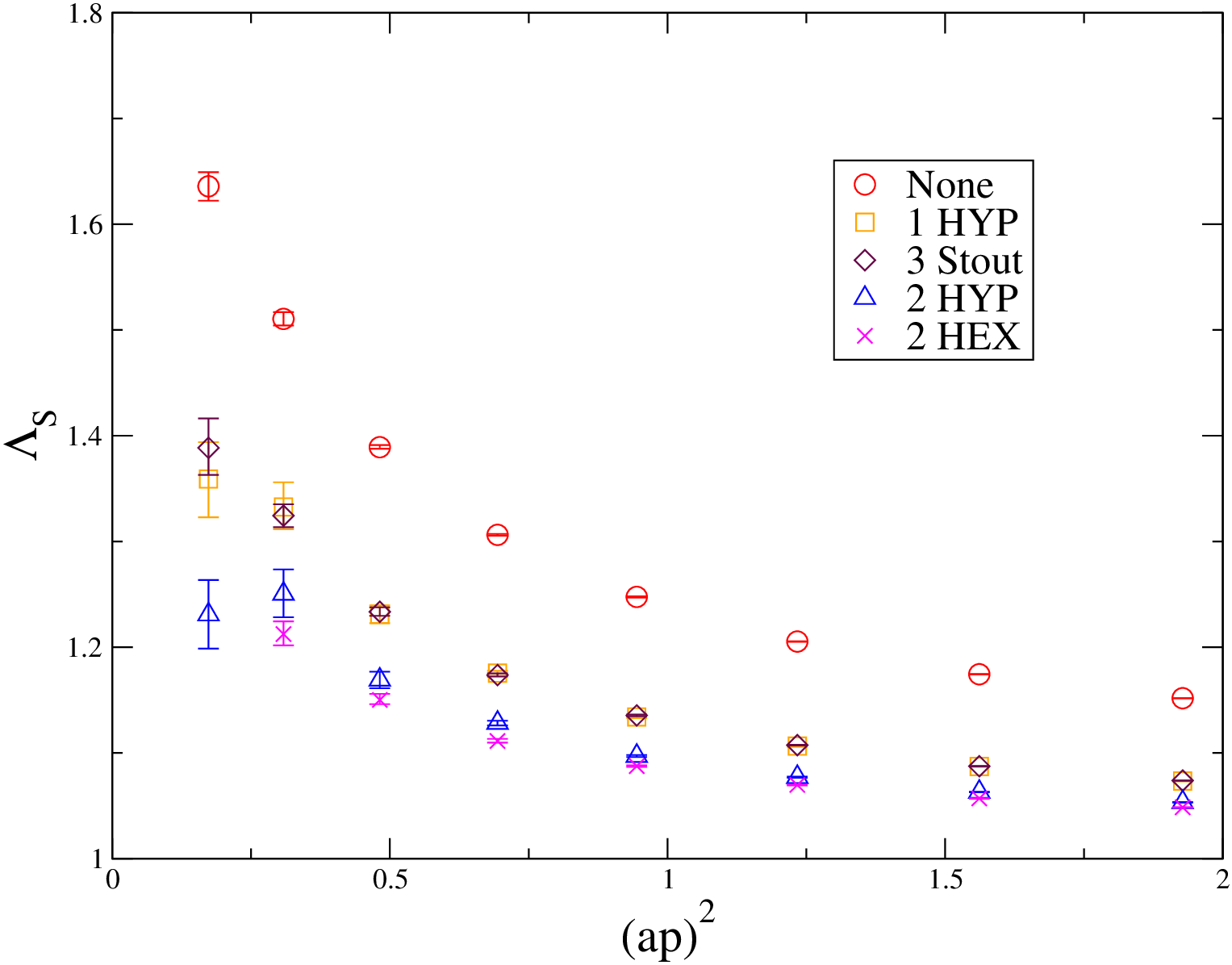}} \\
    \subfloat[$\Lambda_{\cal T}$, $RI-SMOM$]{
\includegraphics*[angle=0,width=0.35\textwidth]{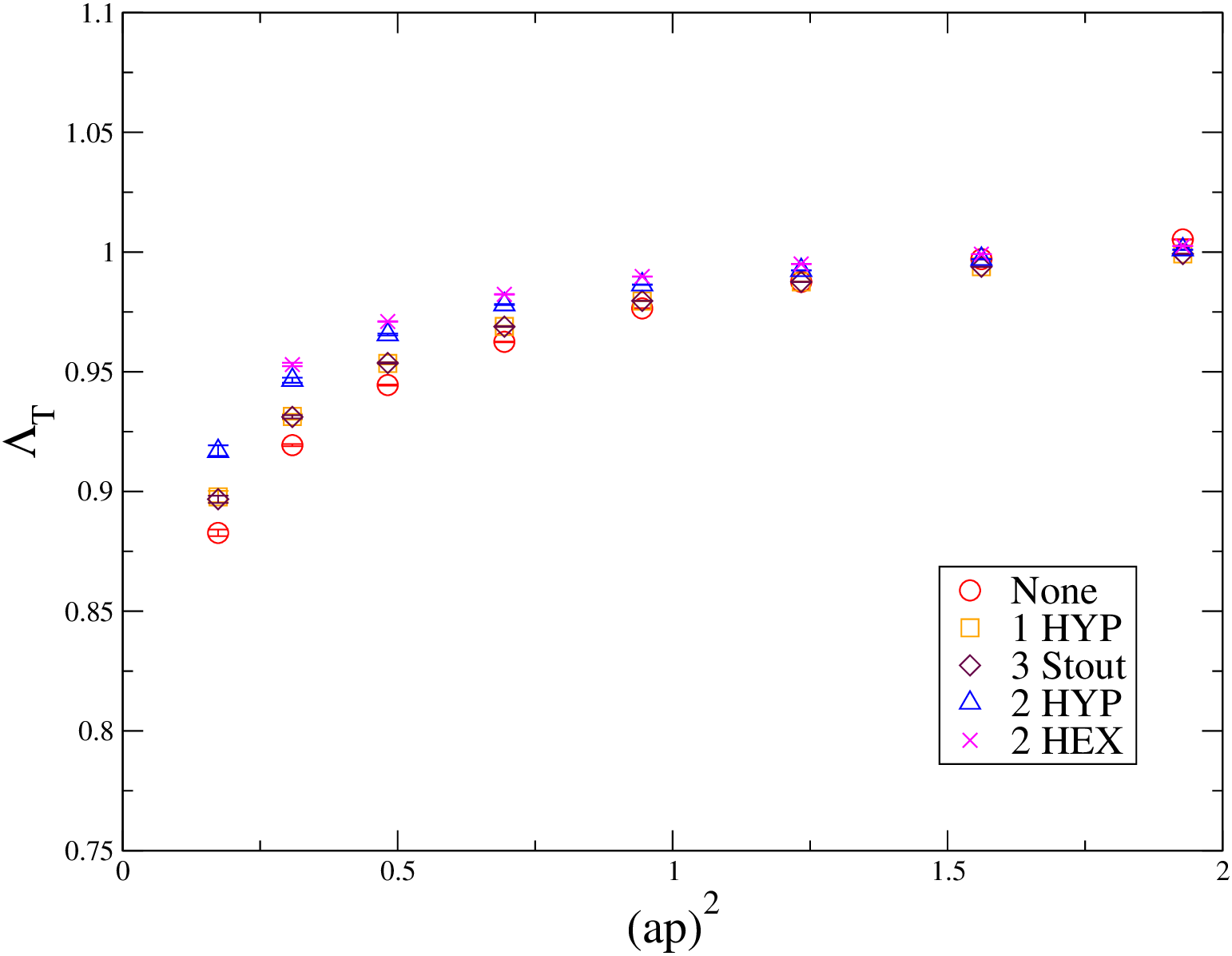}} 
    \subfloat[$\Lambda_{\cal VV+AA}$, $RI-SMOM (\cancel{q},\cancel{q})$]{
\includegraphics*[angle=0,width=0.35\textwidth]{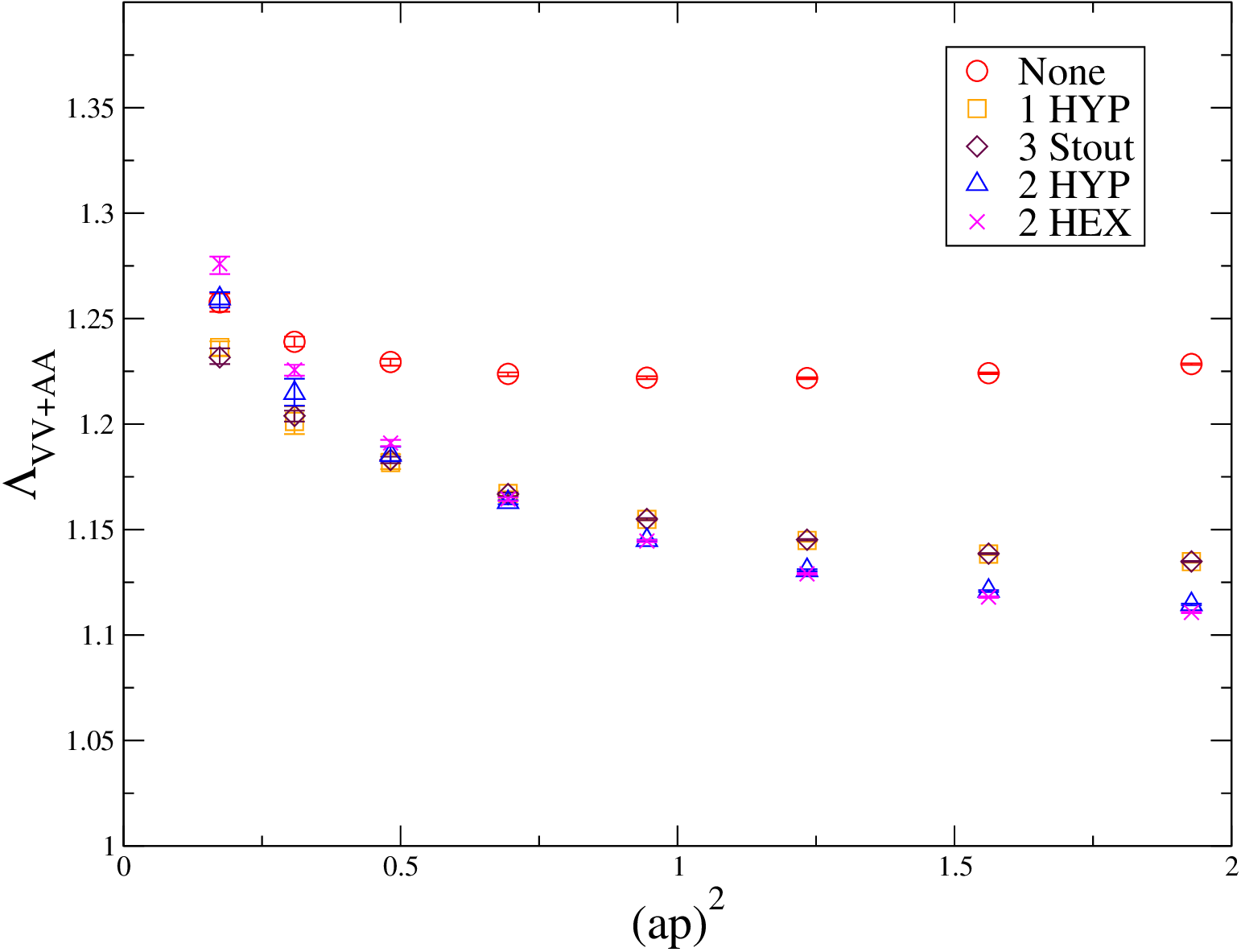}} 
  \end{center}
  \caption{Amputated projected vertex functions in the RI-SMOM scheme, see \cite{Sturm:2009kb}
  for details of this scheme and the relevant projectors. The smearing used is indicated in the
  figure labels. Smearing is seen to drive the renormalization constants closer to
  their tree level values and thus the projected vertex is driven closer to one.
  }
  \label{fig:32004data}
\end{figure}

In Figure \ref{fig:32004data} we show the vertex functions calculated on the $32^3$ lattice with
valence and sea quark masses equal to $0.004$ in lattice units. We are showing this ensemble, 
as it is closest to the chiral and continuum limit, but other ensembles show similar results. The
momentum definition $p_\mu = \frac{2\pi n_\mu + \pi \theta_\mu}{L_\mu}$ is used rather than
another possible definition $\bar{p}_\mu=\frac{1}{a} \sin(ap_\mu)$. 
Here $n_\mu$ is a vector of integers and $\theta_\mu$ a vector parallel to $n_\mu$. 
Explicitly, $p_\mu$ is parallel to $(0,1,1,0)$
and $q_\mu$ is parallel to $(-1,0,1,0)$, the twist added to $p_\mu$ is $(0,\theta,\theta,0)$ and
that added to $q_\mu$ is $(-\theta,0,\theta,0)$ where $\theta$ is varied to vary the magnitude of the momentum.
The definition using $\sin$ agrees at low momenta but gives smaller values at higher momentum for
the same Fourier mode $n_\mu$ and twist $\theta_\mu$. 
In Figure \ref{fig:32004data} the unsmeared vertex functions are shown along with the same quantity
computed using four different types of smearing. 
The effect of smearing is to bring the configurations closer to the free field. Thus the renormalization constants
are driven closer to their tree level values and, because of how the projector is defined, the
vertex functions are driven closer to one. 

\subsection{Step Scaling Functions}
The renormalization constant computed on a smeared gauge field should be combined with
its corresponding operator expectation value, also evaluated on a smeared background and that combination
extrapolated to the continuum limit. In the continuum limit all trace
of the lattice formulation, and hence of any smearing prescription, should vanish. To observe this
we take ratios of renormalization constants at different scales and calculate $\sigma( \mu, s \mu)$
in equation (\ref{eq:sig}). The step scaling function $\sigma$ is scheme dependent. For the bilinear
vertices there are two choices of (non-exceptional) scheme, called SMOM and $\text{SMOM}_{\gamma_\mu}$
which can be used. For $B_K$ there are four non-exceptional schemes (see \cite{Sturm:2009kb} and
\cite{Aoki:2010pe} for definitions of the schemes). We show results in the schemes where the unsmeared data
has been found to agree best with perturbation theory. The four quantities and the scheme we use are given in
table \ref{tab_scheme}.

\begin{table}[ht]
  \begin{center}
 \begin{tabular}{|c|c|}
\hline
    Step scaling function & Scheme\\ \hline
  \hline
 $\sigma_{Z_m}$ & SMOM\\ 
 $\sigma_{Z_q}$ & $\text{SMOM}_{\gamma_\mu}$\\ 
 $\sigma_{Z_T}$ & $\text{SMOM}_{\gamma_\mu}$\\ 
 $\sigma_{Z_{B_K}}$ & $\text{SMOM}_{ \cancel{q},\cancel{q} }$\\ 
\hline
\end{tabular}
  \caption{The schemes in which each step scaling function is calculated.}
  \label{tab_scheme}
  \end{center}
\end{table}

No attempt to account for the systematic errors due to spontaneous chiral symmetry
breaking or other sources is made since these are small, \cite{Aoki:2010dy}, and should
affect smeared and unsmeared data to roughly the same degree.

Since we do not have perfectly matched momentum values $p^2$ between the
two available lattices, we first interpolate the lattice data in $(ap)^2
$ and then choose several values of $p^2$ at which the continuum limit
is taken linearly in $a^2$.  We use the ansatz,
\begin{equation}
\frac{c_{-3}}{ (ap)^6 } + \sum_{i=0}^{2} \frac{c_{-i}}{ (ap)^{2i} } + c_1 (ap)^2.
\end{equation}
This functional form is motivated by the $\frac{1}{p^6}$ behaviour observed at low
momenta in non-exceptional vertex functions \cite{Aoki:2007xm} together with the leading finite lattice spacing error
which for the chirally improved fermion formulation we use are $O(a^2)$. There is also
the intrinsic running of the operator itself, which we account for with a polynomial in
$\frac{1}{(ap)}$. With this ansatz we fit only the data that are in the range of momenta we are 
investigating and these fits are used in all that follows.
Chiral extrapolations are performed linearly in $am$ to 
$-am_{res}$ and are very benign \cite{Arthur:2010ht}. Continuum extrapolations are obtained from a linear
`fit' in $a^2$ using two different lattices. Because our fermion action, domain wall, is chiral all
quantities are automatically off-shell $O(a)$ improved. There is also an $O(a m_{res})$ effect
but as $m_{res}$ is of order $10^{-3}$ this can be safely neglected. Using two lattice spacings lets us account for 
the leading $O(a^2)$ error. 
Only $O(a^4)$ and higher terms remain untreated, which should be small for sufficiently
low $(ap)^2$, lying within the Rome-Southampton scaling window.
This means that continuum extrapolations with domain wall fermions, or other chirally improved actions,
using a smaller number of lattice spacings can be as robust as using more lattice spacings but
an unimproved fermion formulation since the leading non-removed error is the critical quantity.
The detailed range of this window may, however,
depend on the smearing prescription, and if $O(a^4)$ and higher terms become significant they would
manifest themselves as incorrect removal of lattice artefacts in our simple $a^2$
continuum extrapolation.

\begin{figure}[htp]
  \begin{center}
   \centering	
    \subfloat[$\sigma_{Z_m}(1.4\text{ GeV}, \mu)$, no smearing]{
\includegraphics*[angle=0,width=0.35\textwidth]{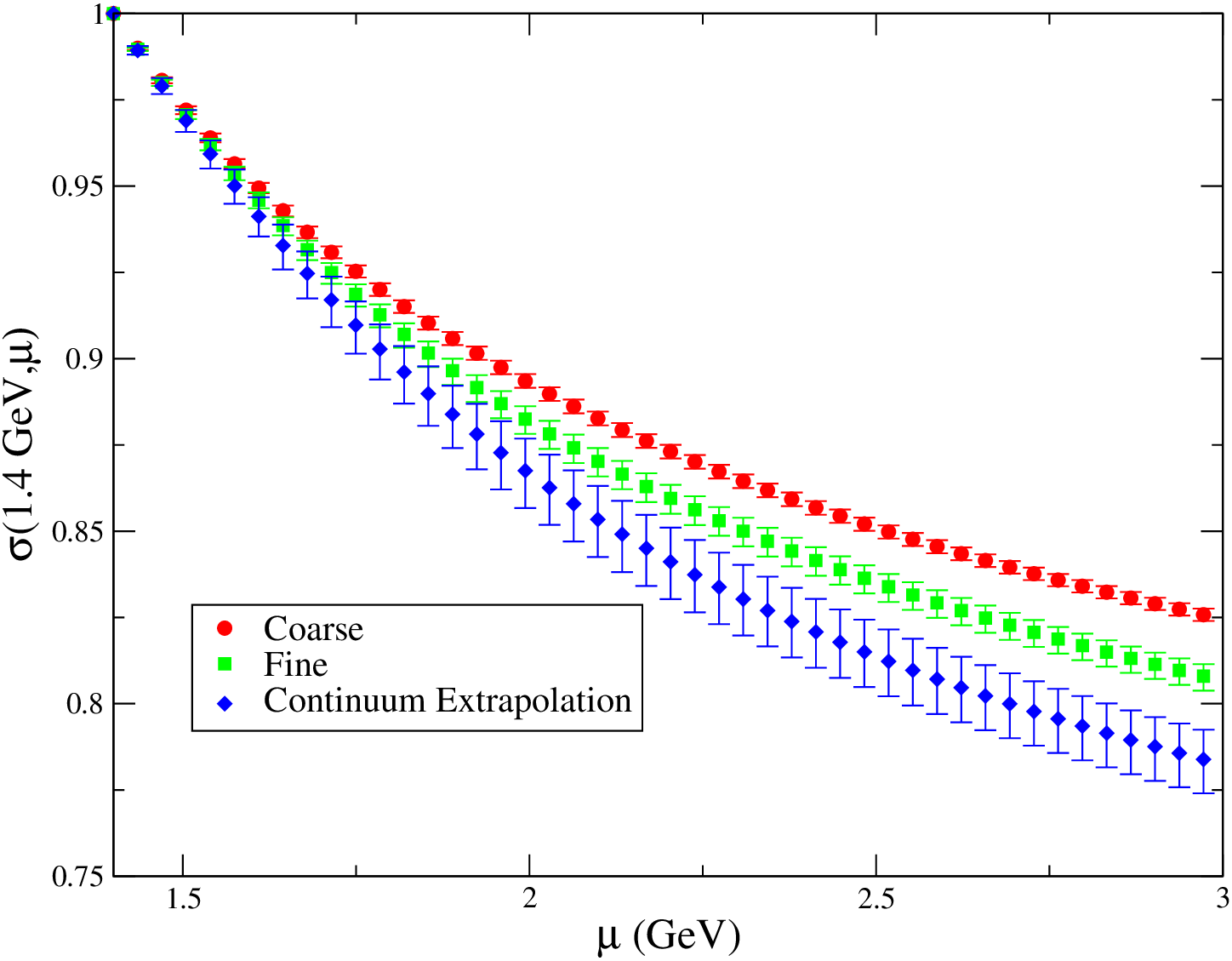}} 
  \qquad
    \subfloat[$\sigma_{Z_m}(1.4\text{ GeV}, \mu)$, 3 stout]{
\includegraphics*[angle=0,width=0.35\textwidth]{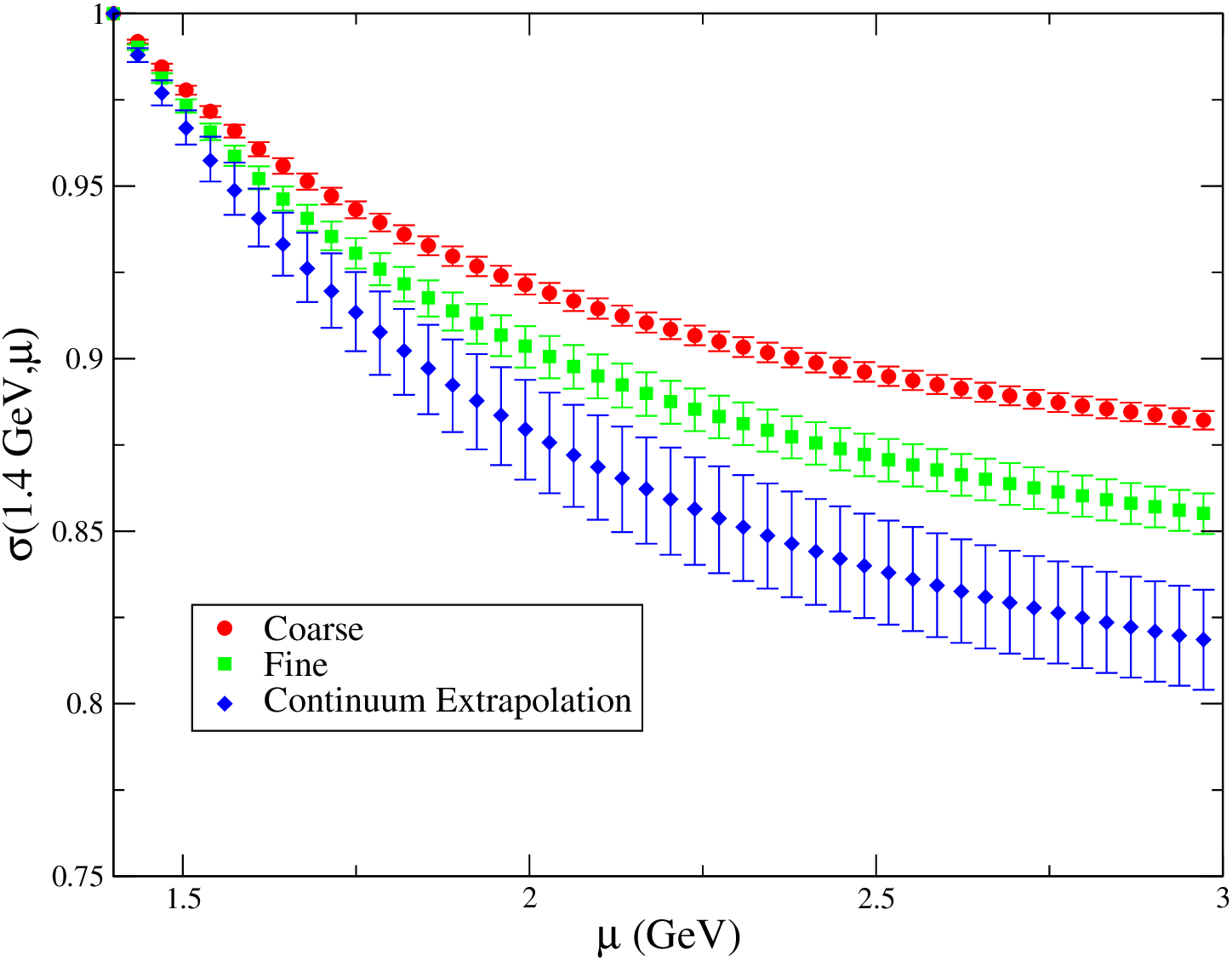}} \\
    \subfloat[$\sigma_{Z_m}(1.4\text{ GeV}, \mu)$, 2 HEX]{
\includegraphics*[angle=0,width=0.35\textwidth]{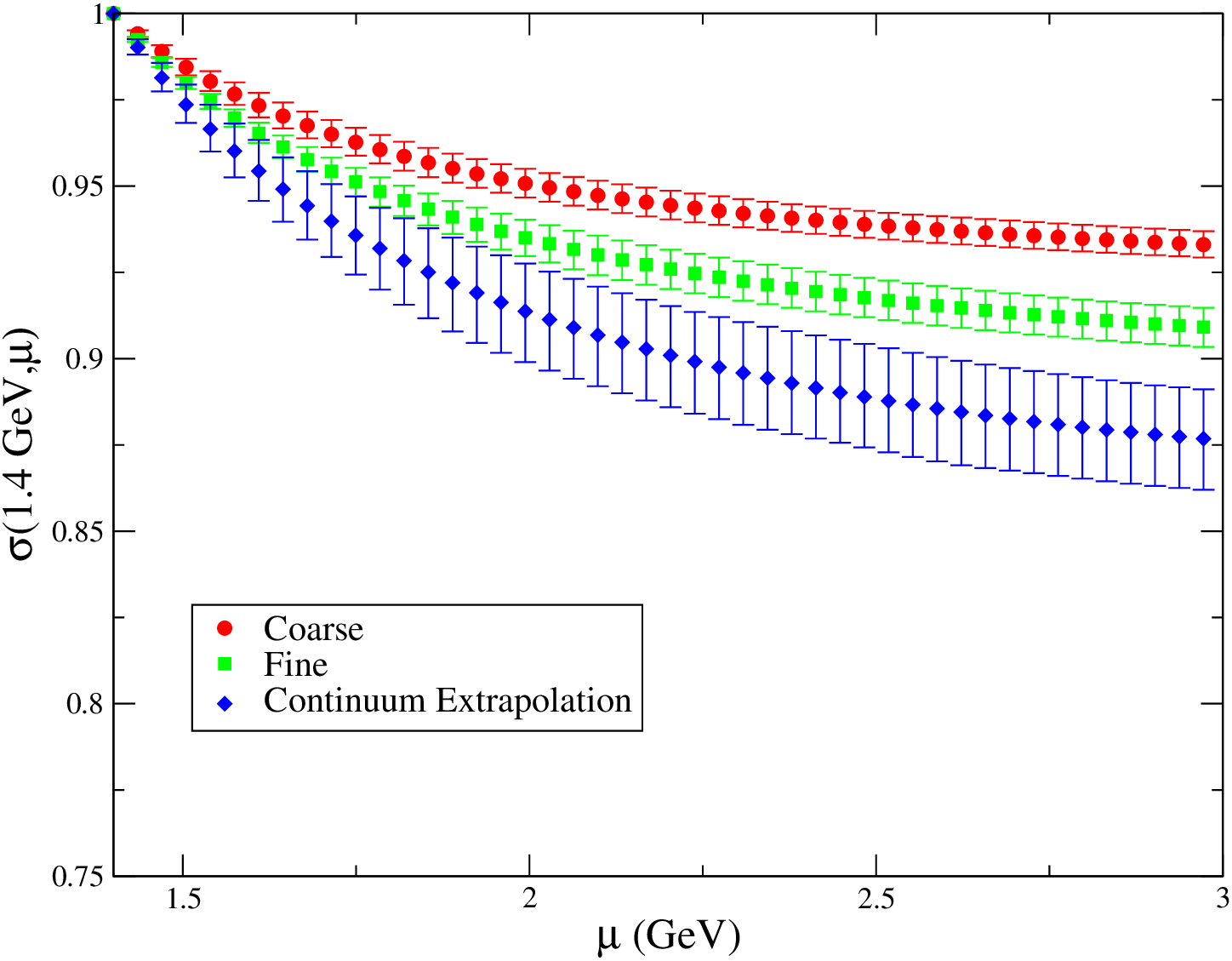}}
  \end{center}
  \caption{The step scaling function for $Z_m$ extrapolated to the continuum in the SMOM scheme.
  (a) in unsmeared (b) uses 3 steps of stout smearing (c) uses 2 steps of HEX smearing. 
  }
  \label{fig:Zmsmear}
\end{figure}

We have chosen to focus on the two differentiable
varieties of smearing, Stout and HEX, because of their usefulness in HMC.
We plot the step scaling function of the mass from a fixed low scale, $1.4 \text{ GeV}$,
to a high scale that varies in the range $(1.4 \text{ GeV}, 3.0 \text{ GeV})$ in figure \ref{fig:Zmsmear}.
The value $1.4 \text{ GeV}$ is chosen to match the choice of scale in \cite{:2012opa}.
In each panel the red data is the step scaling function on the coarse lattice,
the green data is on the fine lattice and the blue data uses these two to extrapolate
in $a^2$ to vanishing lattice spacing. Each of the points in the figure is an arbitrary
scale we have interpolated to and performed the extrapolation at. The different panels show the
three different smearing prescriptions (a) no smearing, (b) 3 hits of stout smearing
and (c) 2 hits of HEX smearing. It is evident that the anomalous running at finite lattice spacing
for the smeared is much weaker than in the unsmeared case. The smearing irons out
the ultraviolet fluctuations leading to a prediction of little to no anomalous running at higher momenta.
It is in this sense that we find it is perhaps more accurate to describe smeared gauge fields as closer to
the trivial configuration than as closer to continuum QCD.

\begin{figure}[htp]
  \begin{center}
    \subfloat[$\sigma_{Z_m}(1.4\text{ GeV}, \mu)$ Extrapolated]{
\includegraphics*[angle=0,width=0.35\textwidth]{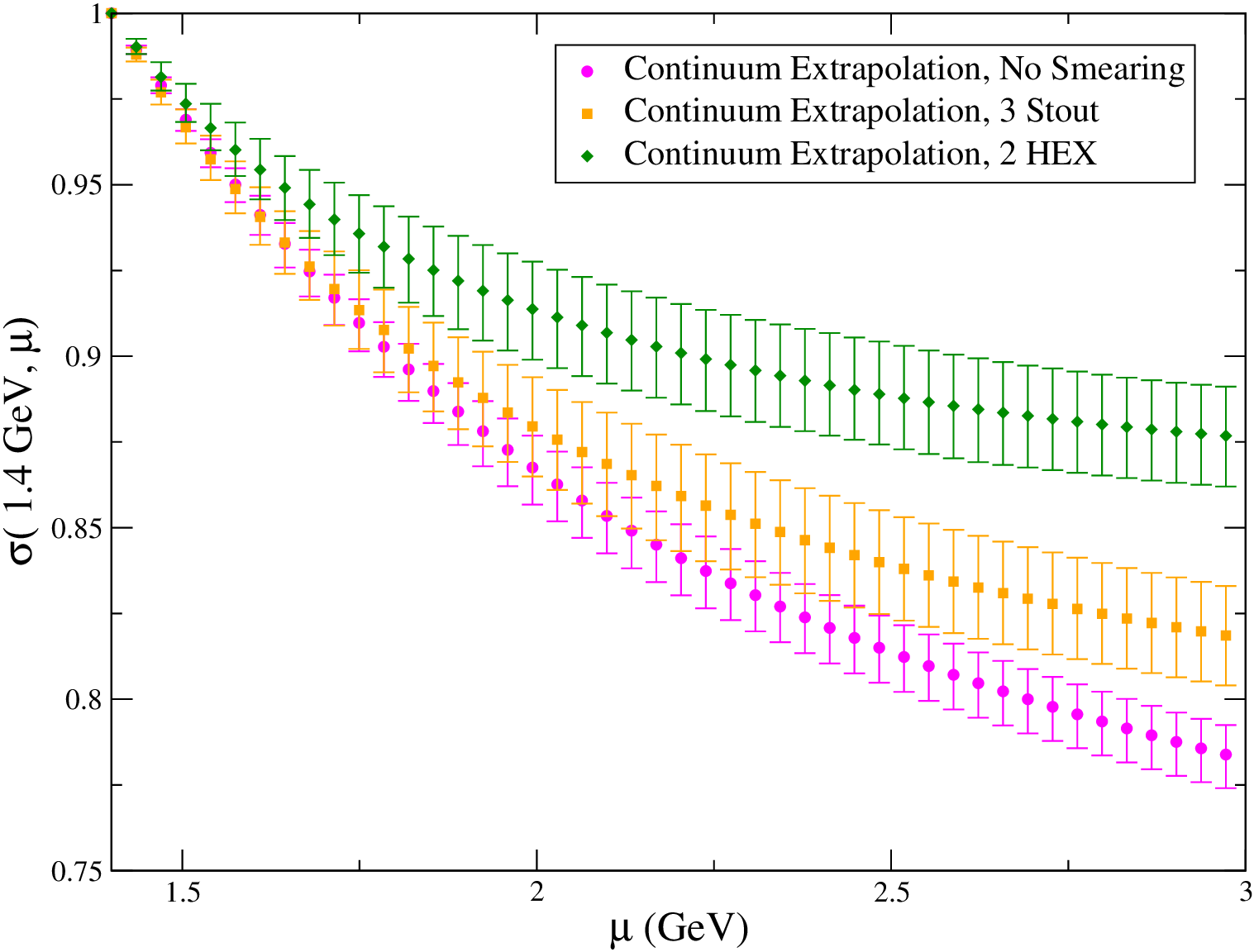}} 
    \subfloat[$\sigma_{Z_q}(1.4\text{ GeV}, \mu)$ Extrapolated]{
\includegraphics*[angle=0,width=0.35\textwidth]{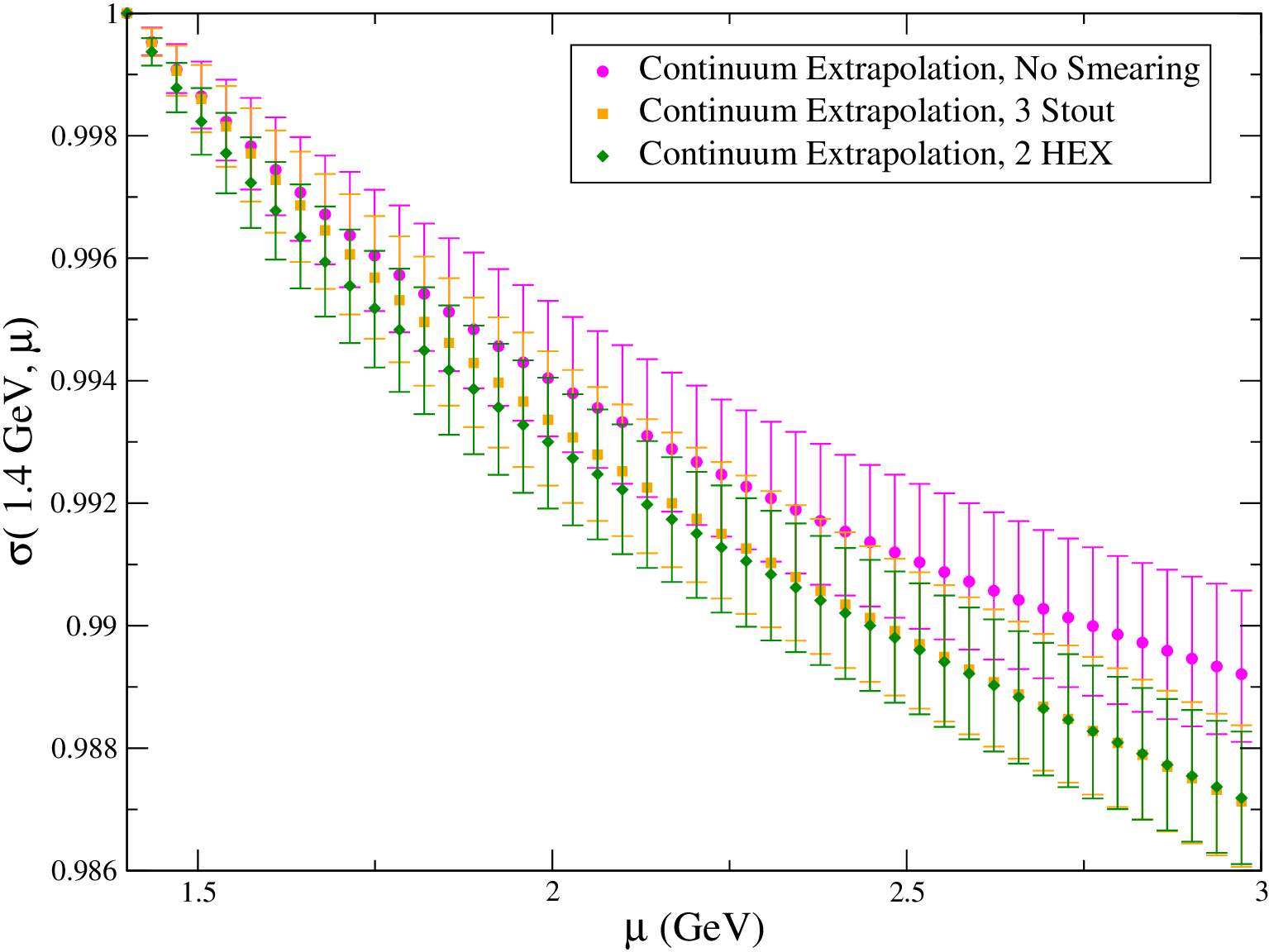}}  \\
    \subfloat[$\sigma_{Z_T}(1.4\text{ GeV}, \mu)$]{
\includegraphics*[angle=0,width=0.35\textwidth]{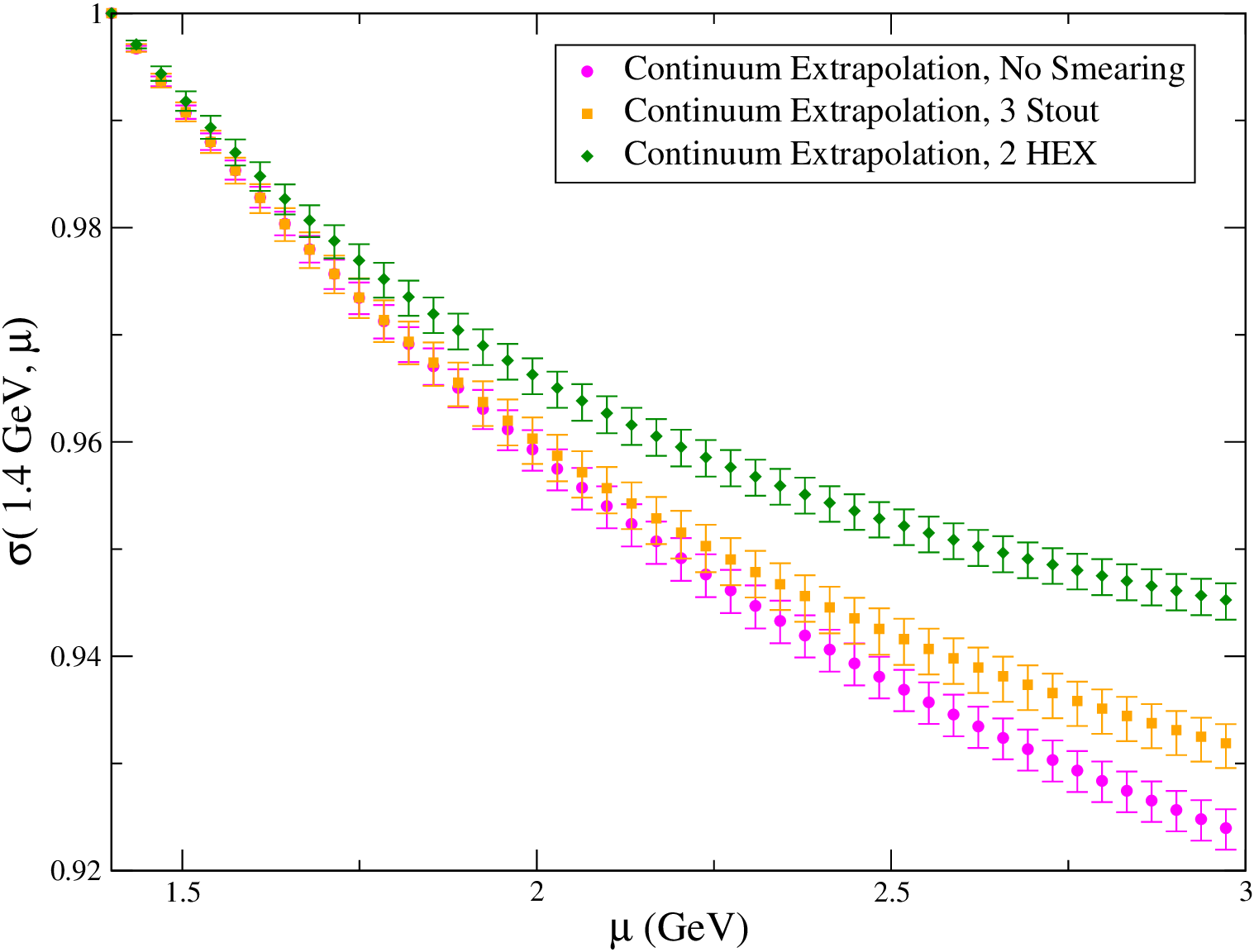}} 
    \subfloat[$\sigma_{Z_{BK}}(1.4\text{ GeV}, \mu)$]{
\includegraphics*[angle=0,width=0.35\textwidth]{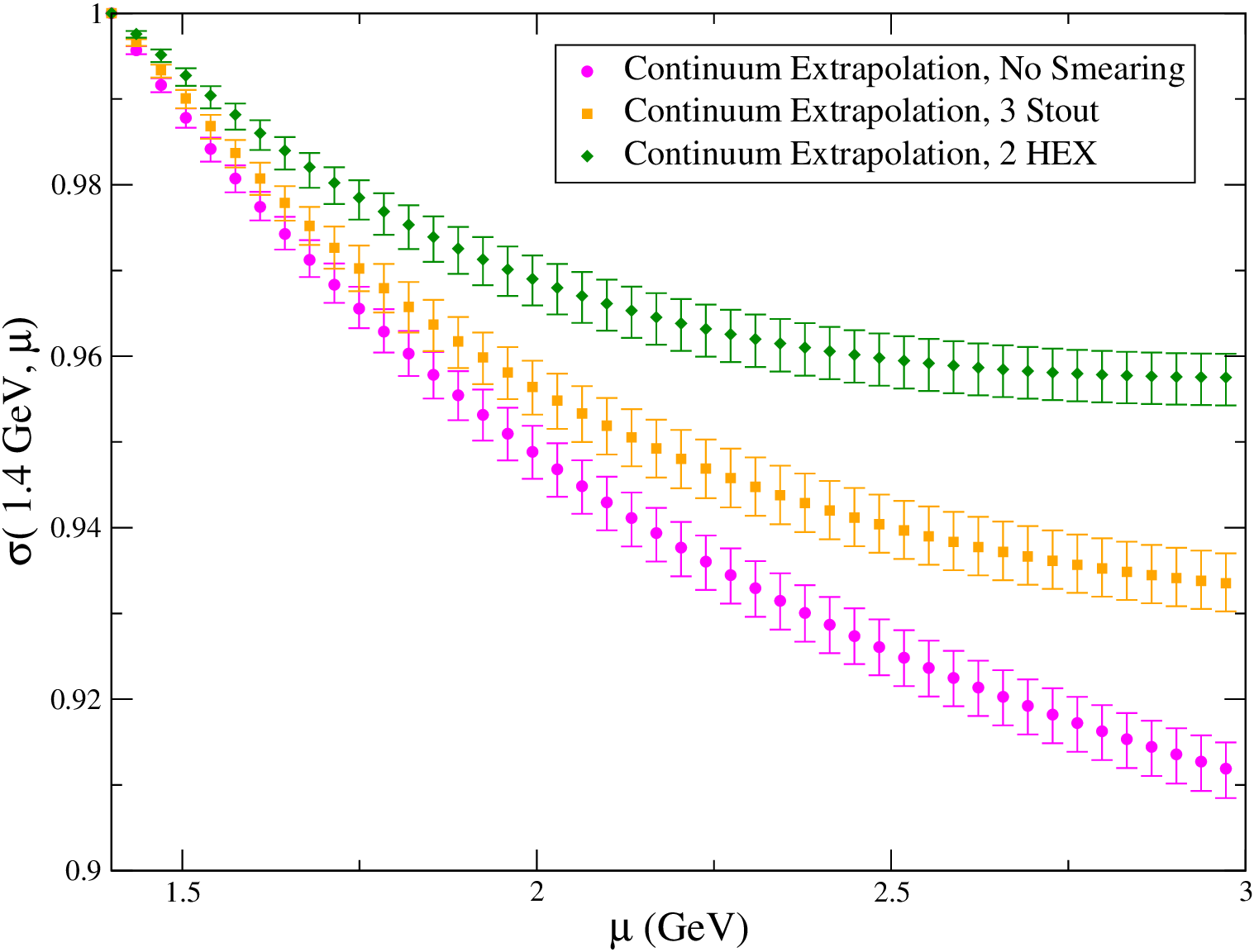}} 
  \end{center}
  \caption{
  The step scaling functions extrapolated to $a^2 = 0$.
  (a) $Z_m$ (b) $Z_q$ (c) $Z_T$ (d) $Z_{BK}$
  }
  \label{fig:SScomp}
\end{figure}

This is not necessarily a problem {\it per se}, providing the effects are well described by only $O(a^2)$ errors
such that we can successfully extrapolate to a universal continuum limit.
If it is the case that we can continuum extrapolate away any difference between the smeared and
unsmeared data then the blue curves in all three figures should be identical.
This can be seen more clearly when all three continuum limits $\sigma_m$ are overlayed in Figure \ref{fig:SScomp}(a).
We also display the same analysis for $\sigma_q$, $\sigma_T$, and $\sigma_{Z_{B_K}}$.
Here, we see a universal slope for the continuum step-scaling function in the region $1.4$ GeV to $1.7$ GeV.
However, we find that the smeared approaches have sufficiently large lattice artefacts to disagree with the thin 
link running in the continuum limit (taken using purely $O(a^2)$ extrapolation) when the physical momenta
exceed around $1.75$ GeV for the HEX smeared case and around $2.0$ GeV for the stout smeared case.
We interpret this as indicating that radiative corrections in smeared link simulations possess a second, hidden, lattice 
spacing defined by the smearing radius that can pinch the Rome-Southampton scaling window compared to thin link simulations.

\begin{figure}[htp]
\psfrag{Perturbation Theory}{\Huge Continuum}
  \begin{center}
    \subfloat[$\gamma_{Z_m}( \mu)$ Extrapolated]{
\includegraphics*[angle=0,width=0.45\textwidth]{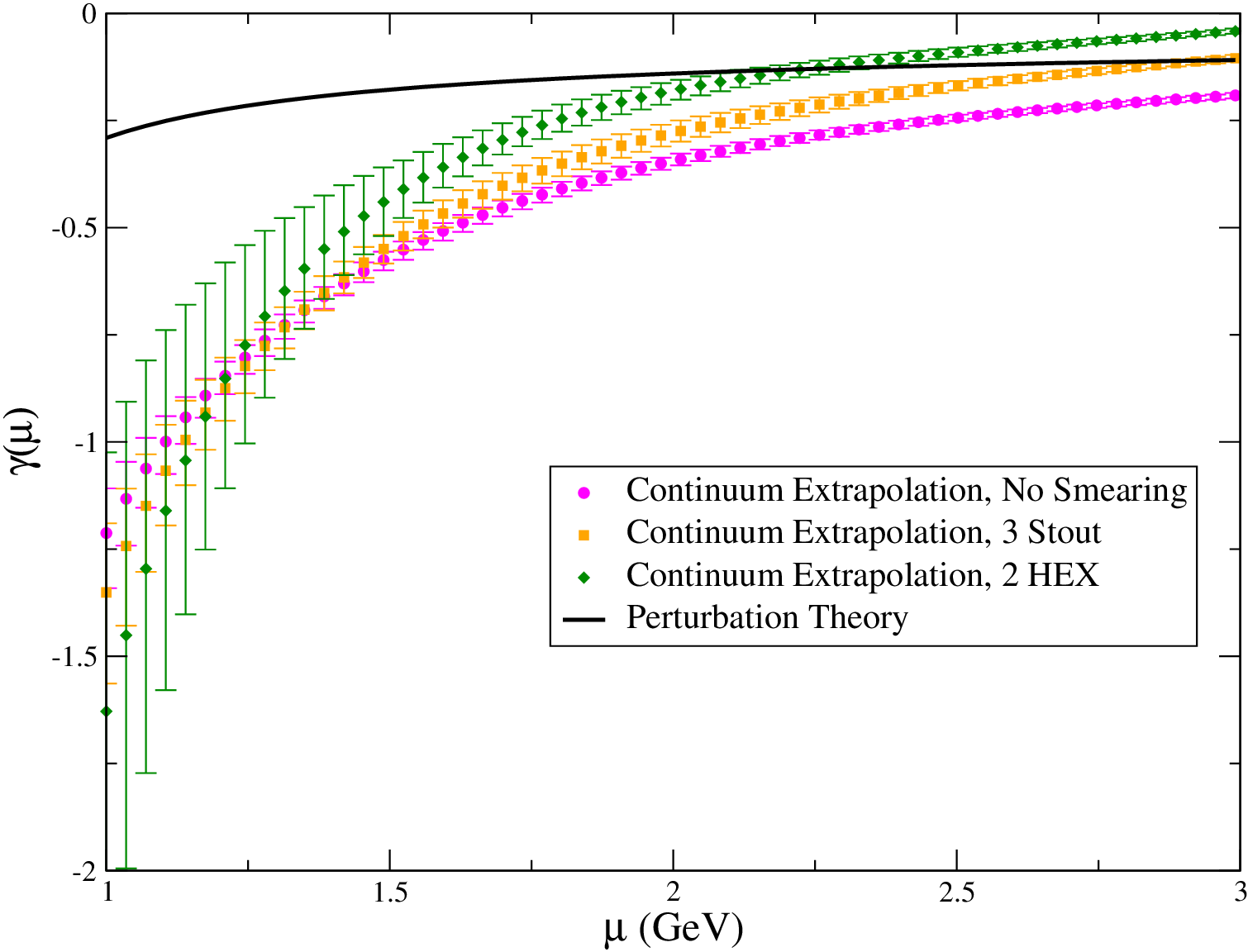}} 
    \subfloat[$\gamma_{Z_q}(\mu)$ Extrapolated]{
\includegraphics*[angle=0,width=0.45\textwidth]{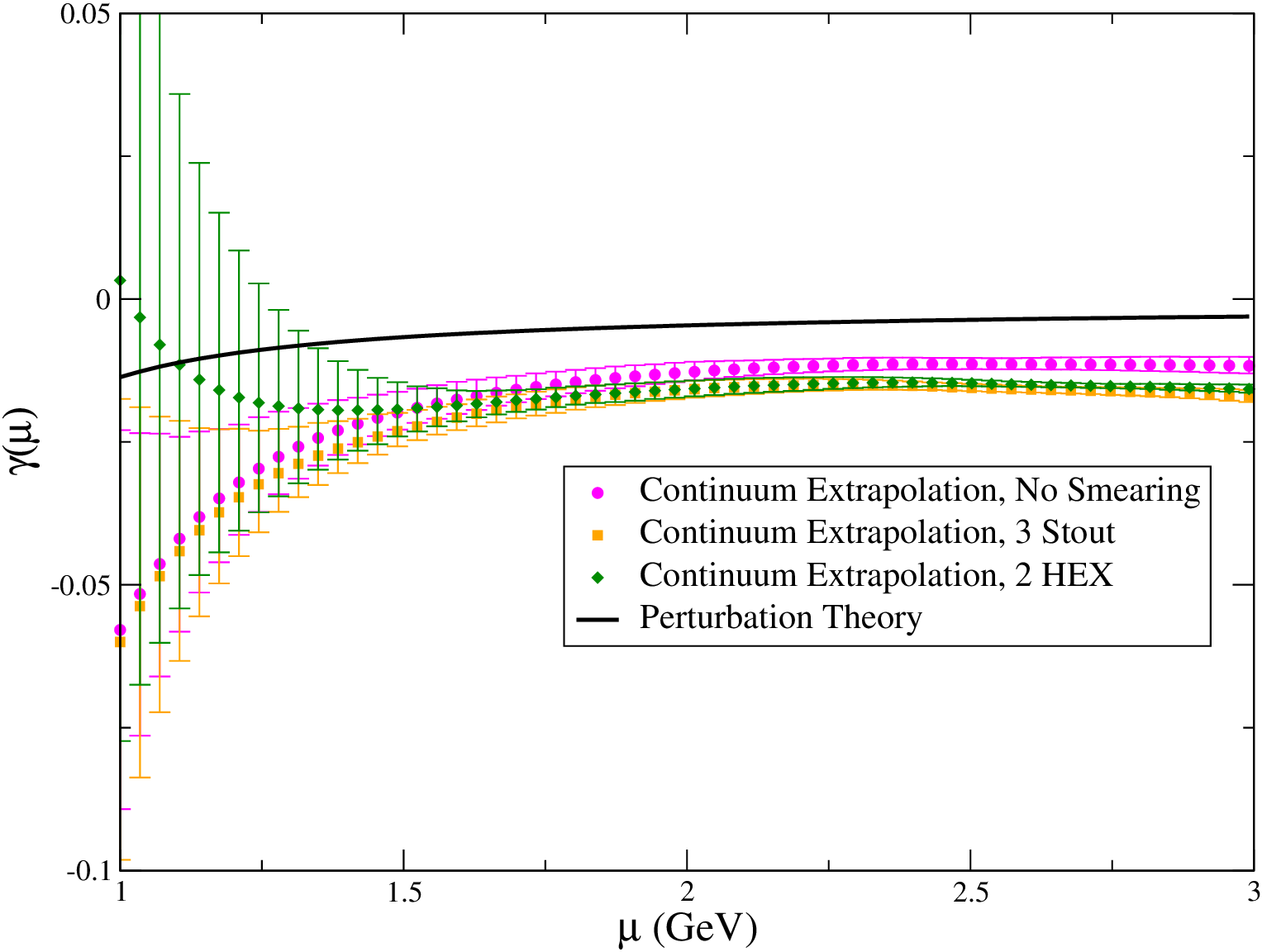}} \\
    \subfloat[$\gamma_{Z_T}(\mu)$ Extrapolated]{
\includegraphics*[angle=0,width=0.45\textwidth]{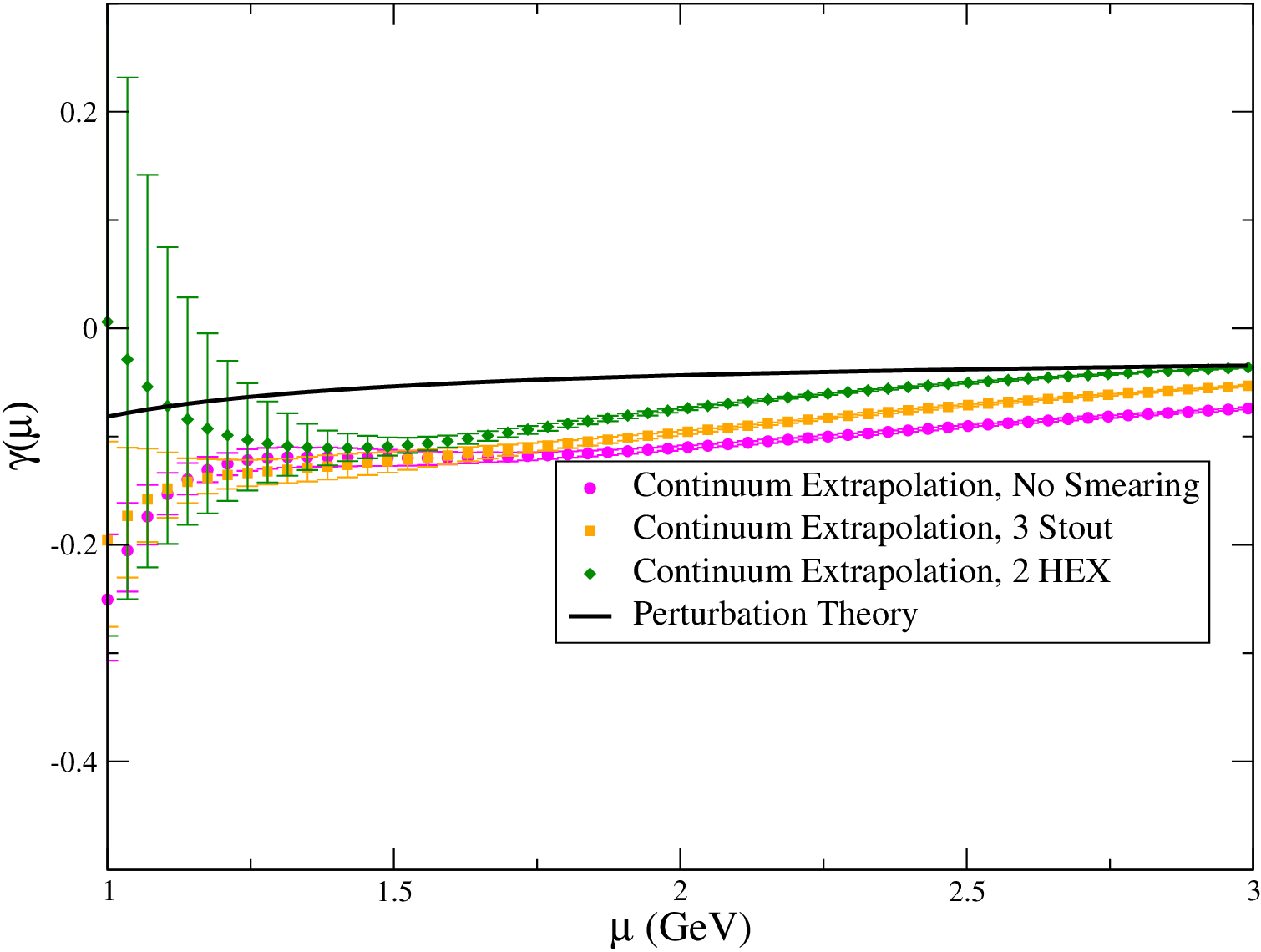}}
    \subfloat[$\gamma_{Z_{BK}}(\mu)$ Extrapolated]{
\includegraphics*[angle=0,width=0.45\textwidth]{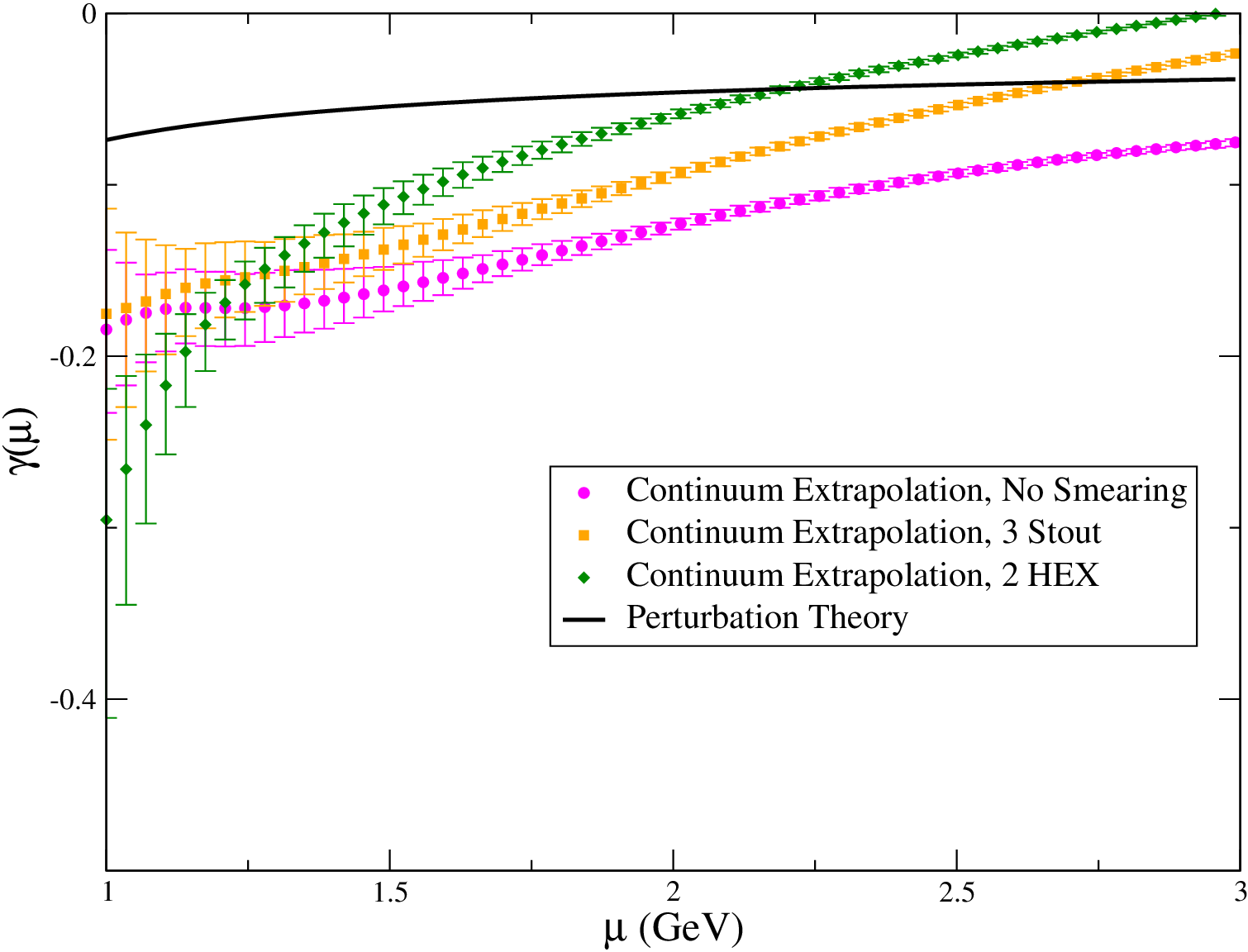}} 
  \end{center}
  \caption{
  The $a^2$ extrapolated anomalous dimension of (a)$Z_m$, (b) $Z_q$, (c) $Z_m$ and (d) $Z_{B_K}$. With three
  smearing prescriptions, none, 3 Stout and 2 HEX overlayed.
  }
  \label{fig:Zgammacomp}
\end{figure}

In Figure \ref{fig:Zgammacomp}(a) we show the anomalous dimension of the mass from $1 \text{ GeV}$ up to $3 \text{ GeV}$
using two different smearings compared to unsmeared data. The anomalous dimension
plots show the effects of smearing more clearly,
up to a scale of the order $1.75 \text{ GeV}$ all three anomalous dimensions agree
with each other. After this the three start to diverge strongly. The solid black line is the
perturbative result for this scheme,
	\begin{equation}
		\gamma_{O}(\mu) = -\sum_i \gamma_{O,i} \left( \frac{\alpha_s(\mu) }{4 \pi} \right)^i
	\end{equation}
where the co-efficients for $\gamma_{m}$ and $\gamma_{q}$ are obtained from \cite{Almeida:2010ns}
those for $\gamma_{T}$ from \cite{Sturm:2009kb} and for $\gamma_{BK}$ from \cite{Aoki:2010pe} and 
$\alpha_s$ is in the $\overline{MS}$ scheme. The highest number of loops available
(two for $B_K$ and the tensor, three for the mass and quark field) is used for the perturbative line. For
$\alpha_s$ we use $\alpha_s(m_Z) = 0.1184$ and run down to $m_b$ with the five flavour expression, to
$m_c$ with the four flavour before running to the desired scale using the three flavour $\beta$ function.
The running uses the four loop $\overline{MS}$ perturbative $\beta$ function. See Table \ref{tab:adcoeff} for the
numerical values of the anomalous dimension coefficients.

Comparing the perturbation theory with the lattice data only the
unsmeared data seems to be asymptotically approaching the perturbative result at high energy,
the smeared data overshoots and badly disagrees. The size of the disagreement between the anomalous
dimensions does not affect the step scaling plots as much because in this region the
absolute value of the anomalous dimension itself is smaller, only when integrated over a sizable energy 
range does this effect become apparent, for example the last point $\sigma_{Z_m}(1.4 \text{ GeV}, 3 \text{ GeV})$
is clearly different depending on the smearing prescription. Table \ref{tab:ssvals} gives the values of
this step scaling function with the different smearings and a significant (up to 10\%) difference is observed
for this quantity, which happens to be the worst we have studied. Calculations
performed using smeared gauge fields will give final results at high scales significantly
different than with unsmeared gauge fields, the difference is larger than the sub percent
errors commonly expected for NPR and should be accounted for or avoided by staying at low scales
($\ll a^{-1}$).

\begin{table}[ht]
  \begin{center}
 \begin{tabular}{|c|c|c|c|}
\hline
     & $\gamma_1$ & $\gamma_2$ & $\gamma_3$ \\ \hline
  \hline
 $Z_m$ & 4 & 66.4763 & 1030.17 \\
 $Z_q$ & 0 & 6.33333 & 99.5426 \\
 $Z_T$ & 1.33333 & 23.6932 &   \\
 $Z_{BK}$ & 2 & 1.26007 &  \\ 
\hline
\end{tabular}
 \begin{tabular}{|c|c|c|}
  \hline
  $\alpha_s(1 \text{GeV})$ & $\alpha_s(2 \text{GeV})$ & $\alpha_s(3 \text{GeV})$ \\ \hline
  \hline
  0.480637 & 0.295752 & 0.245273 \\
\hline
\end{tabular}
  \caption{Coefficients of the anomalous dimension in the RI/SMOM schemes used in this work.
  Numerical values of the $\overline{MS}$ coupling are also listed.}
  \label{tab:adcoeff}
  \end{center}
\end{table}

\begin{table}[ht]
  \begin{center}
 \begin{tabular}{|c|c|c|c|}
\hline
    & Unsmeared & 3 Stout & 2 HEX \\ \hline
  \hline
 $\sigma_{Z_m}$ & 0.782(9) & 0.817(15) & 0.876(15) \\ 
 $\sigma_{Z_q}$ &  0.9891(12) &  0.9869(12) & 0.9870(10) \\ 
 $\sigma_{Z_T}$ & 0.9231(19) & 0.9313(20) & 0.9448(17) \\ 
 $\sigma_{Z_{BK}}$ & 0.9111(32) & 0.9332(34) & 0.9575(30) \\ 
\hline
\end{tabular}
  \caption{The step scaling functions from $1.4$ to $3.0$ GeV, for four different
quantities using two types of smearing compared to the unsmeared. For large steps
the differences can be large.}
  \label{tab:ssvals}
  \end{center}
\end{table}

The plots in Figure \ref{fig:Zgammacomp} show the anomalous dimensions for the quark mass, field, the 
tensor and $B_K$ renormalization. In all cases a similar story is observed. Up to some low scale
the continuum limits all agree, as required for a universal continuum limit. 
However, upon going higher they disagree indicating lattice artefacts beyond $O(a^2)$. From the anomalous dimension plots
the unsmeared data seems to follow the perturbative prediction more closely. 

\section{Conclusions}\label{sec:Conclusions}

The principle
observation is that an $O(a^2)$ extrapolation will not restore agreement between smeared and unsmeared 
data when it includes lattice momenta that are sufficiently high to resolve the smearing radius. Naturally,
if a sufficiently fine lattice were used at a fixed physical momentum universality would be restored.

The inverse lattice spacing for the coarse lattice is approximately $1.75\text{ GeV}$ \cite{Allton:2008pn} which is about the scale at which 
the difference between the stout and unsmeared cases begins to emerge. The stronger smearing, HEX, 
begins to disagree sooner although the exact scale at which there is a disagreement
depends on the operator. The implication of our data is that the upper end of the Rome Southampton window is
being lowered by the smearing. The smearing introduces a new scale into the problem, lower
than the original lattice cut-off, and extrapolating at scales above that with a low order
$a^2$ ansatz will yield incorrect results. 

Comparing
the anomalous dimensions it seems that the unsmeared data is asymptotically approaching perturbation theory, and is quite close
already in the range of momenta we have used.
The smeared data agrees with the unsmeared up to a certain scale, as required for a universal continuum limit,
but disagrees for larger momenta within the range of our data. The unsmeared data also overshoots the perturbative anomalous dimension,  and does not appear to be asymptotically
joining the curve. This is almost certainly the result of a pinched Rome-Southampton window introducing lattice artefacts
not described by our $O(a^2)$ extrapolation. Were sufficiently fine lattices used with these physical scales we expect that
the smeared data would then approach the unsmeared and asymptotically approach the perturbative curve.

This implies that there is a highest lattice momentum
we can use for NPR which depends on the smearing type and is lower for larger smearing radii. Although strictly
speaking we do not know that the unsmeared calculation is lattice artefact free at all scales considered, we can use this as
a comparison point and look at the scale at which the smeared and unsmeared begin to disagree, this also depends 
on the operator and the size of the errors so that being quantitative is difficult. At most we can say with
confidence that this scale is lower than about 2 GeV in physical units and smaller for 2 HEX than 3 Stout with our combination of 
$1.7$ and $2.3$ GeV lattice spacings. 

Smearing the gauge fields has a significant cost benefit for Wilson and chiral fermions, and is becoming
ubiquitous in lattice calculations. However when combined with non-perturbative
renormalization there are some pitfalls. 

We can illustrate this by considering how large an error would have been introduced by using smeared link step
scaling functions in place of thin link for a recent RBC-UKQCD calculation.
Step scaling functions,
computed on the lattices we have used for this work, have been used by RBC/UKQCD
to raise the renormalization scale in previous work, \cite{:2012opa}. In this paper
a coarse lattice spacing was used $a^{-1} = 1.37(1) \text{ GeV}$ meaning that for
NPR only low scales were safely accessible. However using two finer lattices step scaling functions
were produced to raise the renormalization scale up to $3 \text{ GeV}$ thereby reducing
the perturbative conversion error significantly. The $a^{-1} = 1.37(1) \text{ GeV}$ lattice
in this paper used a different action than the lattices used to calculate the step scaling functions
however universality of the continuum limit was used to argue that combining two separate continuum limits
introduced errors of $O(a^4)$.

The step scaling function for the mass calculated in reference \cite{:2012opa} was $0.797(8)$, see table
XVII in this work, reproduced within errors here in table \ref{tab:ssvals}, (we used different configurations
and a different fitting procedure compared to \cite{:2012opa}). The $\overline{MS}$ quark masses calculated in
\cite{:2012opa} were
\begin{equation}
m_{ud}^{\overline{MS}}(3 \text{ GeV}) = 3.05(10) \text{ MeV} \,\,\,\, m_s^{\overline{MS}}(3 \text{ GeV}) = 83.5(2.0) \text{ MeV}.
\end{equation} 
Using $\frac{\sigma_{Zm}^{\text{stout3}}(1.4, 3)}{\sigma_{Zm}^{\text{none}}(1.4, 3)} = 1.025$ and
$\frac{\sigma_{Zm}^{\text{hex2}}(1.4, 3)}{\sigma_{Zm}^{\text{none}}(1.4, 3)} = 1.099$ would change 
the quark masses significantly more than the indicated error:
\begin{equation}
m_{ud}^{\overline{MS}}(3 \text{ GeV}, \text{3 Stout}) = 3.13(10) \text{ MeV} \,\,\,\, m_s^{\overline{MS}}(3 \text{ GeV}, \text{3 Stout}) = 85.6(2.1) \text{ MeV}
\end{equation}
\begin{equation}
m_{ud}^{\overline{MS}}(3 \text{ GeV}, \text{2 HEX}) = 3.35(11) \text{ MeV} \,\,\,\, m_s^{\overline{MS}}(3 \text{ GeV}, \text{2 HEX}) = 91.8(2.2) \text{ MeV}.
\end{equation}
Now that calculations
of key parameters such as quark masses and $B_K$ have become so precise the effect of smearing on NPR can
be a dominant systematic error if not properly controlled.
Proper control appears to require keeping the lattice momenta $(ap)^2\le 1$ for all vertex functions
included in the renormalisation due to the pinched Rome-Southampton window, while we found that unsmeared
data can make use of somewhat larger lattice momenta. We also found that the effects of stout smearing
(at 2.5\%) were much smaller than the effects of HEX smearing  (at 10\%). Both of these could have been avoided with 
more conservative lattice momenta, but this would have prevented the successful connection with perturbation theory
at 3 GeV.

The observed trends 
in the reduced anomalous dimension at high energy as well as the flattening of the step scaling functions and 
the renormalization constants moving closer to one
are a strong indication, though not a proof, that smearing reduces the effective
lattice cut-off. 

We note that the effect of reduced perturbative corrections has been
loosely described as showing that smearing gauge fields brings the theory
closer to the continuum limit.
In this paper we have presented clear evidence based on smeared fermionic vertex functions that such 
gauge fields possess a second, hidden, coarse lattice 
spacing associated with the smearing radius, and introduce
a narrower Rome-Southampton scaling window.
In light of the removal of the UV content of the gauge field, 
and the results of this paper, we find it is perhaps more 
accurate to describe smeared gauge fields as closer to the trivial configuration in a 
momentum region where QCD has real dynamics to describe, particularly if we seek to 
run, non-perturbatively, into the perturbative regime to connect to perturbation theory.

Renormalizing at lower scales is not a problem in principle for lattice calculations.
A problem arises when performing a perturbative conversion to the $\overline{MS}$
scheme. The perturbative expansions do not converge very well at low energy
and raising the scale by even $1\text{ GeV}$ can dramatically reduce the perturbative
systematic error.  However, this is expensive to do by brute force. 
As such, combining Rome-Southampton renormalization at low scales
with some safe, non-perturbative running to high scales before converting may be necessary for high precision
calculations of renormalization constants on smeared backgrounds. 

If future ensembles are generated using smeared links then step scaling functions \cite{Arthur:2010ht} may be useful
for reducing systematic errors in renormalization calculations by allowing us to always operate within a safe
momentum range.

\section*{Acknowledgements}
Simulations made use of the STFC funded DiRAC facility.
We acknowledge support from the following grants ST/K005790/1, 
ST/K005804/1, ST/K000411/1, ST/H008845/1, ST/J000329/1.
The CP$^3$-Origins centre is partially funded by the Danish National 
Research Foundation, grant number DNRF90.

\end{document}